\documentclass[
 reprint,
superscriptaddress,
floatfix,
 amsmath,amssymb,
 aps,
]{revtex4-2}
\usepackage{graphicx}

\usepackage[T1]{fontenc}

\usepackage{dcolumn}
\usepackage{bm}
\usepackage{braket}
\usepackage[dvipsnames]{xcolor}
\usepackage{soul}
\usepackage{gensymb}
\usepackage{afterpage}
\usepackage{hyperref}


\newcommand{\mytitle}{Microscopic Understanding of Thermal-magnon Transport in Low-damping Ferrimagnetic Thin Films}

\begin{document}
\title{\mytitle}

\author{Lerato Takana}
\affiliation{Department of Applied Physics, Stanford University, Palo Alto, CA, 94305, USA}
\affiliation{Geballe Laboratory for Advanced Materials, Stanford University, Stanford, California 94305, USA}

\author{Katya Mikhailova}
\affiliation{Department of Applied Physics, Stanford University, Palo Alto, CA, 94305, USA}
\affiliation{Geballe Laboratory for Advanced Materials, Stanford University, Stanford, California 94305, USA}

\author{Junwei Tong}
\affiliation{Department of Physics , University of Austin, Austin, TX, 78712}

\author{Xiangcheng Liu}
\affiliation{Department of Physics , University of Austin, Austin, TX, 78712}

\author{Kwangyul Hu}
\affiliation{Department of Physics and Astronomy, University of Iowa, Iowa City, Iowa, 52242}

\author{Juan A. Hofer}
\affiliation{Department of Physics and Center for Advanced Nanoscience, University of California, San Diego}

\author{Guanxiong Qu}
\affiliation{Department of Physics and Astronomy, University of California, Irvine, Irvine, CA 92697}

\author{Clare C. Yu}
\affiliation{Department of Physics and Astronomy, University of California, Irvine, Irvine, CA 92697}

\author{Ivan K. Schuller}
\affiliation{University of California, San Diego}

\author{Michael Flatt{\'e}}
\affiliation{Department of Physics and Astronomy, University of Iowa, Iowa City, Iowa, 52242}

\author{Xiaoqin Li}
\affiliation{Department of Physics , University of Austin, Austin, TX, 78712}

\author{Yuri Suzuki}
\affiliation{Department of Applied Physics, Stanford University, Palo Alto, CA, 94305, USA}
\affiliation{Geballe Laboratory for Advanced Materials, Stanford University, Stanford, California 94305, USA}
\affiliation{Stanford Institute for Materials and Energy Sciences, SLAC National Accelerator Laboratory, Menlo Park, California 94025, USA}

\date{\today}
\begin{abstract}
Thermally generated magnons enable heat-driven spin transport in magnetic insulators, yet the relative importance of multiple microscopic mechanisms governing their propagation remains incompletely understood. Here, we investigate thermal magnon transport in low-damping Li$_{0.5}$Al$_{1.0}$Fe$_{1.5}$O$_4$/Pt nanodevices using a nonlocal spin Seebeck geometry that separates magnon transport from local thermoelectric effects. Thermal imaging establishes a detector region outside the thermal healing length, enabling intrinsic nonlocal measurements. We find that thermal magnon transport is strongly suppressed by magnetic fields far above saturation, and further that thermal magnon transport decreases with increasing temperature despite an increasing magnon population. Brillouin light scattering reveals the key microscopic mechanism driving this effect: increasing field reduces the group velocity of backward volume magnons, directly reducing the magnon spin diffusion length.  Micromagnetic simulations reproduce this behavior only when a temperature-dependent exchange stiffness is included. These results identify magnon group velocity and exchange stiffness as key parameters governing thermal magnon transport in ferrimagnetic thin films.
\end{abstract}

\maketitle

Pure spin currents in magnetic insulators offer a promising platform for information transfer because spin-wave transport can dissipate substantially less energy than charge-based electronics \cite{demokritov2012magnonics,hoffmann2015opportunities,Moradi2019,Parkin2015}. Magnon currents can be generated by electromagnetic radiation, spin torques, optical excitation, acoustic waves, and thermal gradients \cite{uchida2010spin,gomez2020differences,chumak2015magnon,giles2017thermally}. Because optical and electrical excitation inevitably produce heating, thermally generated magnons often coexist with magnons generated by other mechanisms and are therefore an ubiquitous component of magnon transport experiments. However, the microscopic transport of thermal magnons remains incompletely understood. In particular, magnon dispersion, group velocity, scattering, and exchange stiffness all collectively determine the propagation of thermally populated magnons, but their relative roles remain unresolved. Addressing these questions is essential for understanding heat-driven spin transport and optimizing material properties to develop spintronic technologies which convert waste heat into useful spin electromotive force.

Thermal magnon transport is commonly studied via the spin Seebeck effect (SSE) \cite{uchida2010spin}, where a temperature gradient drives a thermal magnon current that is detected electrically in an adjacent nonmagnetic metal via the inverse spin Hall effect (ISHE). However, if the detecting metal is placed within the thermal healing length (an area with significant temperature gradient) of the magnetic insulator, the same temperature gradient can also generate thermoelectric voltages \cite{gao2022magnon} in the metal, complicating the interpretation of the SSE measurements \cite{meyer2017observation,gomez2020differences,cornelissen2016magnetic,gao2022magnon,cornelissen2017nonlocal,giles2017thermally,oyanagi2020magnetic,sola2015evaluation,vlietstra2014simultaneous,cosset2024nonreciprocal,weiler2012local,taghinejad2025low}. Establishing the spatial regime in which thermal gradients are well defined is therefore essential for isolating intrinsic thermal magnon transport. 

Recent studies further indicate that thermal magnon transport is highly sensitive to intrinsic magnetic properties. In particular, magnon non-conserving processes in vanadium tetracyanoethylene (in a longitudinal SSE geometry) and anisotropy-induced suppression of thermal transport in yttrium iron garnet (YIG) (in a nonlocal geometry) highlight the importance of scattering and magnetic anisotropy in governing thermal magnon propagation \cite{kurfman2026magnon,taghinejad2025low}. In contrast to magnon scattering, the role of the real part of the magnon energy (dispersion) in nonlocal transport remains less clear. Spin-wave studies in YIG have shown that the field dependence of the group velocity depends strongly on the spin-wave mode and magnetic configuration: field-dependent group velocities have been observed for Damon–Eshbach (DE) modes \cite{qin2018propagating,krysztofik2017characterization}, whereas recent measurements of forward-volume spin waves (FVSW) in perpendicularly magnetized YIG found an approximately field-independent group velocity \cite{shiota2026perpendicular}. Whether and how such mode-dependent dispersion effects govern the propagation of thermally populated magnons in a nonlocal geometry remains unexplored. This motivates controlled measurements of the magnon dispersion and transport in magnetic insulators in a nonlocal geometry.

Here we investigate the microscopic mechanisms governing thermal magnon transport in Li$_{0.5}$Al$_{1.0}$Fe$_{1.5}$O$_4$ (LAFO)/Pt bilayers by combining nonlocal spin Seebeck measurements with infrared thermometry, Brillouin light scattering (BLS) spectroscopy, and micromagnetic simulations of thermal magnon transport. Infrared thermal imaging directly determines the spatial temperature profile, allowing us to identify a transport regime in which the detector lies outside the thermal healing length and the measured nonlocal signal reflects intrinsic magnon transport. Within this regime, we show that the magnetic-field dependence of the thermal magnon signal originates from the field dependence of the magnon spin diffusion length. Brillouin light scattering measurements of the backward volume (BV) magnon dispersion directly reveal a reduction in magnon group velocity with increasing magnetic field, providing the microscopic origin of the reduced diffusion length and suppressed thermal magnon transport. We further demonstrate that thermal magnon transport decreases with increasing temperature despite the larger thermal magnon population. Micromagnetic simulations of the thermal magnon transport establish the exchange stiffness as the dominant microscopic mechanism controlling the temperature dependence of thermal magnon transport. 

\begin{figure}[!htbp]
\includegraphics[scale = 0.62]{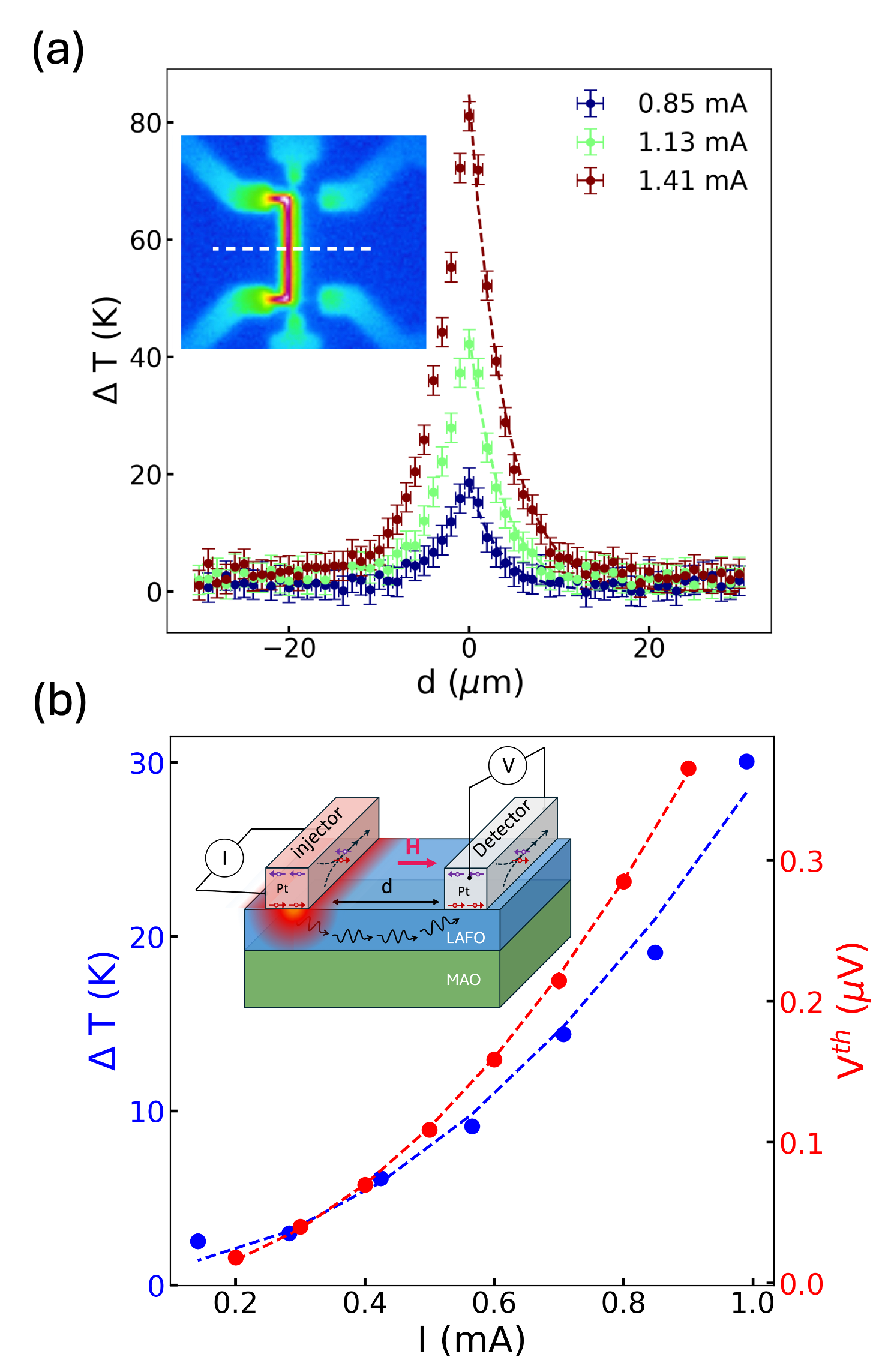}
\caption{(a) Mid-wave infrared (MWIR) thermal imaging of a LAFO/Pt device ($d = 1~\mu$m) under an RMS injector current of 1.4 mA. The inset shows the measured thermal map, and the dashed line indicates the direction of the extracted temperature profiles. The temperature rise above 300 K as function of the distance from the injector, d, is well described by an exponential decay, yielding the thermal healing length used to define the intrinsic nonlocal transport regime. (b) \textit{Left axis:} Temperature rise at the injector as a function of applied current obtained from the MWIR measurements, showing the expected quadratic dependence associated with Joule heating. \textit{Right axis:} Second-harmonic nonlocal thermal voltage measured at 250 K and 0.3 T as a function of injector current, demonstrating that the thermal magnon signal scales quadratically with current and therefore linearly with the temperature gradient. Inset: Schematic of the nonlocal measurement geometry, where Joule heating in the Pt injector generates thermal magnons that are detected in a spatially separated Pt strip via the inverse spin Hall effect (ISHE).}
    
    \label{fig:1}
\end{figure}

We investigate thermal magnon transport in 13 nm thick Li$_{0.5}$Al$_{1.0}$Fe$_{1.5}$O$_4$ (LAFO) films epitaxially grown on single-crystal (001) MgAl$_2$O$_4$ substrates. The films exhibit low magnetic damping providing an ideal platform for studying thermal magnon transport (Supplementary Section I \cite{supp}) \cite{daisy2023aluminum,zheng2023ultra,takana2025low,zheng2020ultra}. In the nonlocal device geometry shown in the inset of Fig.~\ref{fig:1}(b), Joule heating in a Pt injector generates a controlled temperature gradient that drives a thermal magnon spin current in LAFO via the spin Seebeck effect \cite{uchida2010spin}. The magnon spin current propagates to a spatially separated Pt detector, where it is converted into an electrical voltage through the inverse spin Hall effect (ISHE). Thermal magnon transport is isolated from current-induced spin-orbit torque contributions using harmonic lock-in detection, with the second-harmonic voltage, $V^{th}$, corresponding to the thermal magnon signal. We have investigated the dependence of $V^{th}$ on the temperature gradient, magnetic field, and ambient temperature to identify the microscopic mechanisms governing thermal magnon transport.

\begin{figure*}[!htbp]
\includegraphics[width=\textwidth]{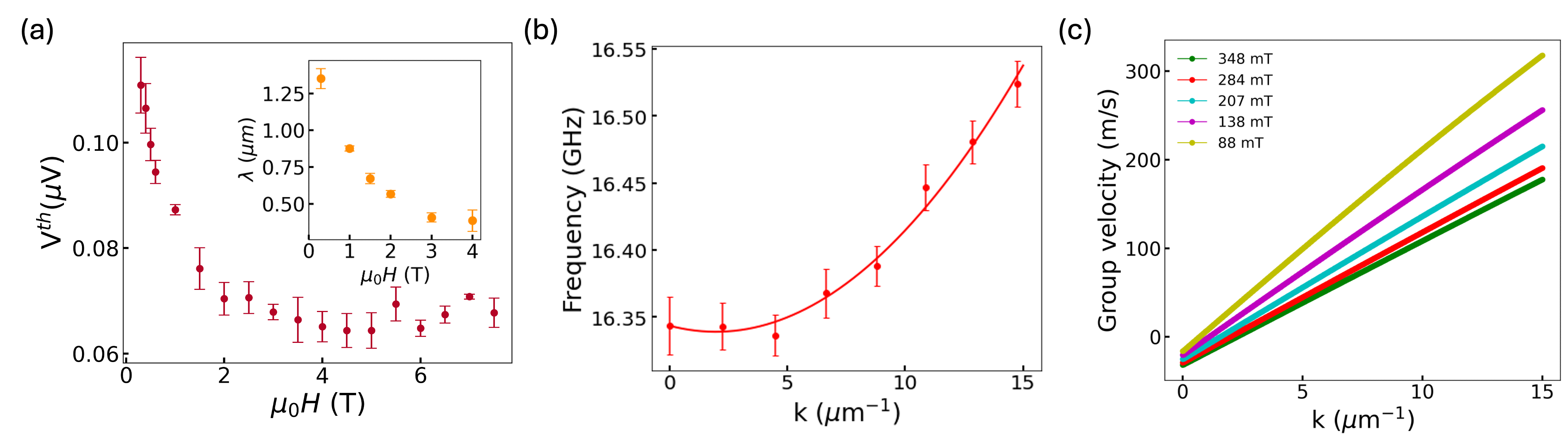}
\caption{Magnetic-field dependence of thermal magnon transport and its microscopic origin. (a) Nonlocal thermal voltage as a function of magnetic field measured at 250 K with an injector current of 0.5 mA. The inset shows the corresponding magnetic-field dependence of the extracted magnon spin diffusion length, demonstrating that the suppression of the thermal signal is accompanied by a reduction in the magnon propagation length. (b) Backward-volume (BV) magnon dispersion of Li$_{0.5}$Al$_{1.0}$Fe$_{1.5}$O$_4$ measured by Brillouin light scattering (BLS) at $\mu_0H = 284$ mT and 300 K. Symbols represent the experimental data and the dashed line is a fit to Eq.~(1). (c) Magnetic-field and wave-vector dependence of the calculated corresponding magnon group velocity, $v_g=2\pi\partial f/\partial k$, extracted from the measured dispersion. The reduction of group velocity with increasing magnetic field provides the microscopic origin of the observed decrease in spin diffusion length and the suppression of thermal magnon transport.}
   
    \label{fig:2}
\end{figure*}

Before analyzing thermal magnon transport, we directly characterize the spatial extent of Joule heating in the LAFO/Pt device using mid-wave infrared (MWIR) thermal imaging \cite{hofer2025mechanically}. As shown in Fig.~\ref{fig:1}(a), RMS (root mean square) currents ranging from 0.85 - 1.4 mA generate a local temperature rise at the injector that decays exponentially, $\Delta T  = \Delta T_{inj}e^{-\frac{d}{\ell_H}}$, along the device. From line profiles of the temperature distribution, we extract an exponential thermal healing length $\ell_H \approx 3.5~\mu$m for all the currents, and determine that $\Delta T \propto \nabla T/\ell_H$ which can be reduced to $\Delta T \propto \nabla T$, since $\ell_H$ is constant in the current range studied (Supplementary Section III \cite{supp}). This establishes a well-defined spatial regime for separating thermal gradients from the detector region. We therefore perform nonlocal measurements with an injector–detector separation of $d = 4~\mu$m, where the detector is outside the region of significant temperature gradient, thereby isolating intrinsic nonlocal magnon transport from local thermoelectric contributions.

We first verify that the measured nonlocal signal originates from thermal magnon generation. A static in-plane magnetic field of 0.3 T, larger than the saturation field of LAFO, is applied perpendicular to the Pt strips to define a uniform magnetization direction and maximize the detected ISHE signal, Fig.~\ref{fig:1}(b) \cite{maekawa2013spin, hoffmann2007pure,daisy2023aluminum,takana2025low,zheng2023ultra}. The measured voltage at the detector scales quadratically with current, $V^{th} \propto I^2$, consistent with Joule heating in the injector. Combined with the MWIR measurements (Supplementary Section III \cite{supp}), this demonstrates that $V^{th}$ is directly proportional to the applied temperature gradient, $V^{th} \propto \nabla T$.

To identify the microscopic origin of thermal magnon transport, we next investigate the magnetic-field dependence of the thermal magnon signal. Measurements were performed at 250 K while sweeping the magnetic field from 0.3 T to 8 T. Throughout this field range, the LAFO magnetization remains saturated, ensuring that changes in the thermal signal do not arise from magnetization rotation. As shown in Fig.~\ref{fig:2}(a), the thermal magnon voltage decreases with increasing magnetic field over a broad field range extending well beyond magnetic saturation before approaching a field-independent value at high fields. Similar field-induced suppression has been reported in YIG and antiferromagnetic insulators \cite{cornelissen2016magnetic,oyanagi2020magnetic,thiery2017spin,taghinejad2025low,gomez2020differences,lebrun2018tunable}, although its microscopic origin remains under debate. 

To determine whether this behavior originates from changes in the transport of individual magnons instead of changes in the population, we extracted the magnon spin diffusion length by measuring the distance dependence of the nonlocal thermal signal over multiple injector-detector separations and fitting the data using the diffusion model described in Supplementary Section IV \cite{supp}. Although measurements at distances beyond the thermal healing length are essential for establishing that the detected signal is dominated by intrinsic magnon transport, the signal-to-noise ratio becomes increasingly limited at large separations. We therefore extract $\lambda$ primarily from devices with $d<5$ $\mu m$, where the signal remains sufficiently large for reliable fitting. The extracted spin diffusion length, shown in the inset of Fig.~\ref{fig:2}(a), exhibits the same two-regime behavior as the thermal voltage: a low-field regime in which $\lambda$ decreases with increasing magnetic field followed by a nearly field-independent regime at higher fields. The close correspondence between the thermal signal and the spin diffusion length indicates that the magnetic-field dependence of thermal magnon transport is governed by changes in the magnon propagation length.

\begin{figure*}[!htbp]
\includegraphics[width=0.8\textwidth]{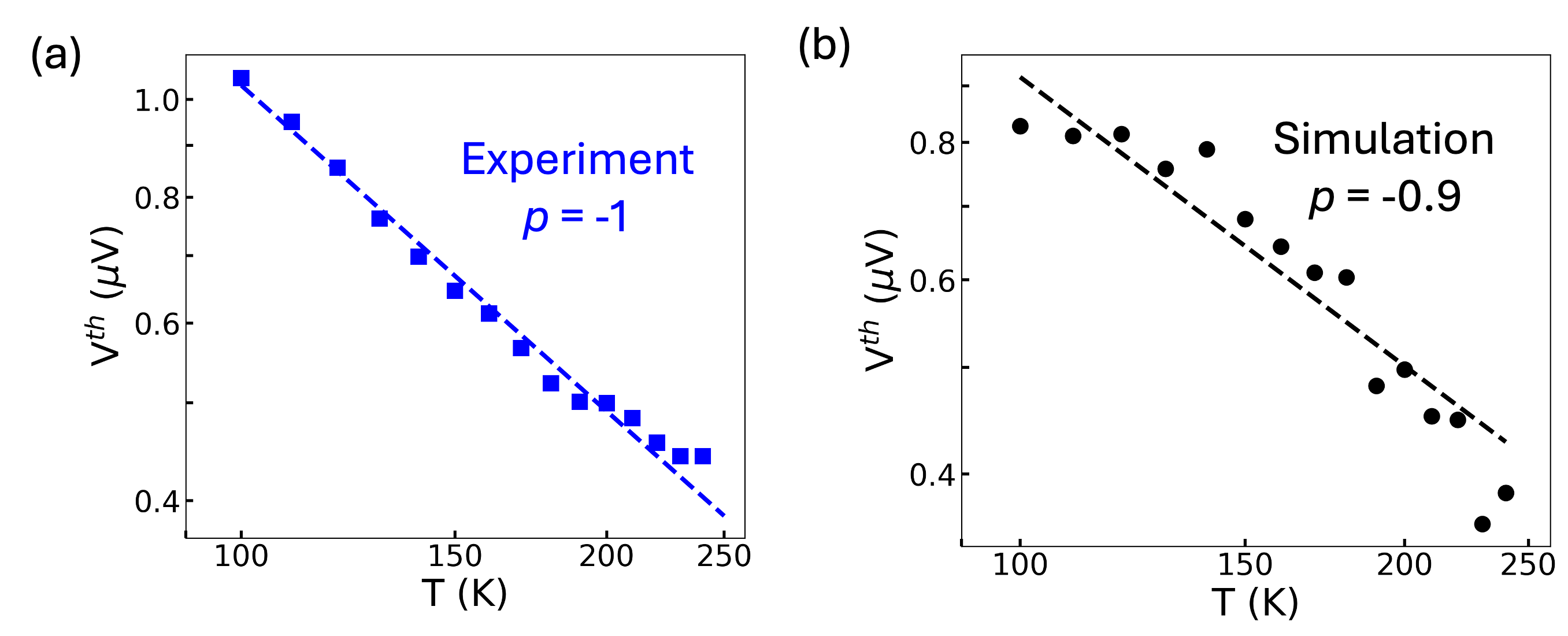}
\caption{Temperature dependence of thermal magnon transport. (a) Thermal magnon voltage measured as a function of temperature at $\mu_0H=0.3$ T and an injector current of 0.5 mA. The data are plotted on logarithmic axes to determine the power-law dependence of the thermal signal on temperature. The dashed line is a power-law fit $V^{th}\propto T^p$ with exponent $p=-1$. (b) Micromagnetic simulations of the thermal magnon signal, plotted on logarithmic axes, for an RMS injector current of 1.4 mA and $\mu_0H=0.3$ T using a temperature-dependent exchange stiffness, $A_{\mathrm{ex}}(T)\propto [M_s(T)/M_s(0)]^2$. The simulated temperature dependence reproduces the experimentally observed power-law behavior, supporting exchange-stiffness renormalization as the dominant origin of the temperature dependence of thermal magnon transport. The dashed line is a power-law fit with exponent $p=-0.9$.}
    \label{fig:3}
\end{figure*}

To understand the microscopic nature of the field-dependence of $\lambda$, we measured the magnon dispersion using Brillouin light scattering (BLS) in the BV geometry (Supplementary Section V \cite{supp}), where the magnetic field is applied parallel to the magnon propagation direction, a geometry consistent with the electrical measurements shown in Fig.~\ref{fig:1}). The in-plane wavevector was tuned via the optical incidence angle, enabling $k$-resolved magnon spectroscopic measurements. We plot the BLS dispersion relation in Fig.~\ref{fig:2}(b) with a fit to the magnon dispersion~\cite{kalinikos1986theory}:
{\small
\begin{multline}
f = \frac{\gamma \mu_0}{2 \pi} \sqrt{(H + H_{4,\parallel}+l_{ex}k_\parallel^2)} \\
\times \sqrt{\left(H  + H_{2,\perp}+H_{4,\parallel} + l_{ex}k_\parallel^2-F(k_\parallel t)M_s+ M_s\right)}
\label{eq:1}
\end{multline}}
where $\gamma$ is the gyromagnetic ratio, $\mu_0$ is the vacuum permeability, $H$ is the applied magnetic field, $H_{4,\parallel}$ is the in-plane four-fold anisotropy field, $H_{2,\perp}$ is the out-of-plane two-fold anisotropy field,  $M_s$ is the saturation magnetization, $l_{ex} \equiv 2 A_{ex}/\mu_0 M_s $ is the exchange field stiffness with $A_{ex}$ being the exchange stiffness, $F(x) = 1 - \frac{1-e^{-|x|}}{|x|}$, $k_\parallel$ is the in-plane wavevector, and $t$ is the film thickness. To fit the data, we used $\mu_0 H_{4,\parallel} = -19$ mT, $\mu_0 H_{2,\perp}=892$ mT and $\mu_0 M_s = 126$ mT \cite{daisy2023aluminum}; an in-plane exchange field stiffness of $\mu_0 l_{ex} = 3131 \pm 442 ~\mathrm{T}\mathring{\mathrm{A}}^2$ was obtained, consistent with prior reports \cite{tong2026direct, daisy2023aluminum}. From Eq. (\ref{eq:1}), we compute the group velocity $v_g = 2\pi \partial f / \partial k$, as shown in Fig.~\ref{fig:2}(c). The fit to the data in Fig.~\ref{fig:2}(b) has a minimum at 2 $\mu$m$^{-1}$ (Supplementary Section V \cite{supp}). The sign change associated with a negative $v_g$ at lower k reflects the BV magnon character \cite{prabhakar2009spin}. Furthermore, Fig.~\ref{fig:2}(c) shows that increasing the magnetic field systematically reduces $v_g$ for the BV mode in LAFO, consistent with previous observations of field-dependent group velocities for DE modes in YIG films \cite{qin2018propagating,krysztofik2017characterization}. Because the magnon diffusion constant,  $D_m \sim v_g^2\tau_c$, scales quadratically with the magnon group velocity, the observed reduction in $v_g$ provides the microscopic origin of the field-dependent reduction in the magnon spin diffusion length and, consequently, the suppression of the nonlocal thermal magnon signal shown in Fig.~\ref{fig:2}(a).

We next investigate the temperature dependence of thermal magnon transport to identify the microscopic origin of its thermal evolution. Measurements were performed for temperatures above 100 K, where phonon-drag \cite{adachi2010gigantic,rodriguez2023dominance} and interfacial contributions \cite{guo2016influence} to the spin Seebeck effect are expected to be negligible. As shown in Fig.~\ref{fig:3}(a), the thermal magnon signal decreases monotonically with increasing temperature despite the increasing thermal magnon population. A similar trend has been reported in YIG \cite{cornelissen2016magnetic,oyanagi2020magnetic,thiery2017spin}; however, its microscopic origin remains unresolved. In the following, we show that the observed temperature dependence is quantitatively explained by the temperature dependence of the exchange stiffness. 

To determine whether the temperature dependence of the exchange stiffness governs thermal magnon transport, we performed micromagnetic simulations using MuMax3 \cite{Vansteenkiste2014MuMax}. The simulations incorporate the experimentally determined temperature dependence of the magnetic parameters (Supplementary Section I \cite{supp})  and assume a power-law dependence of the exchange stiffness, $A_{ex}(T)\propto[M_s(T)/M_s(0)]^n$ \cite{atxitia2010multiscale,Niitsu_2020,Mulazzi_2008}, where $n$ is treated as an effective exponent appropriate for ferrimagnetic LAFO \cite{Moreno2024TempFerri,Hirst2022TempMn2Au,Rozsa2022TempAFerro,Mikuni2025FMRAFerro}. Fig. \ref{fig:3}(b) shows the simulated thermal magnon signal for $I=1$ mA, $\mu_0H=0.3$ T, and $n=2$. Despite fluctuations arising from the stochastic thermal field \cite{Vansteenkiste2014MuMax}, the simulated signal decreases monotonically with increasing temperature and follows a power-law dependence, $V^{th}\propto T^p$, with $p=-0.9$, in good agreement with the experimentally measured exponent of $p=-1$ (Fig. \ref{fig:3}(a)). Together with the failure of simulations using a temperature-independent or weakly temperature-dependent exchange stiffness (Supplementary Section VI \cite{supp}), these results demonstrate that incorporating the temperature dependence of the exchange stiffness is essential for reproducing the observed suppression of the thermal magnon transport.

Our results establish that the magnetic-field dependence of thermal magnon transport in LAFO is governed by the field dependence of the magnon propagation dynamics. Previous studies of spin-wave propagation have shown that the magnetic-field dependence of the group velocity depends on the magnon mode and magnetic configuration \cite{qin2018propagating,krysztofik2017characterization}. However, the role of this mode-dependent group velocity in nonlocal thermal magnon transport has remained unclear. Previous models of thermal magnon transport in yttrium iron garnet, terbium iron garnet, and other magnetic insulators have considered field-dependent diffusion constants, magnon populations, and relaxation processes to describe the field dependence of the nonlocal signal \cite{gao2022magnon,cornelissen2016magnetic,kikkawa2015critical,jin2015effect,rezende2014magnon,xiao2010theory,mihalceanu2018temperature,kurfman2026magnon}. By combining nonlocal transport measurements with direct measurements of the magnon dispersion, we show that the suppression of thermal magnon transport in LAFO is accompanied by a reduction of the magnon spin diffusion length, while BLS measurements directly reveal a corresponding reduction in the group velocity of the thermally relevant BV modes with increasing magnetic field. Because the diffusion constant scales as $D_m \sim v_g^2\tau_c$, the field-induced reduction in group velocity provides a direct microscopic mechanism for the reduced magnon propagation length and the suppressed thermal magnon signal.

The temperature dependence reveals a distinct microscopic mechanism. Although the thermal magnon population increases with temperature, the nonlocal thermal signal decreases because magnon transport becomes progressively less efficient. Within the diffusive picture, $D_m \sim v_g^2\tau_c$, changes in both magnon propagation velocity and magnon-conserving relaxation time can strongly influence the transport length. Our micromagnetic simulations reproduce the observed temperature dependence only when the exchange stiffness is allowed to decrease with temperature. This temperature-dependent exchange stiffness modifies the magnon dispersion and reduces the magnon propagation velocity, while enhanced magnon–magnon and magnon–phonon scattering further limit the transport. The agreement between experiment and simulation therefore demonstrates that the temperature evolution of the exchange interaction is a central microscopic ingredient governing thermal magnon transport in LAFO, rather than the thermal magnon population alone.

Together, these results establish the underlying mechanism for thermal magnon transport in ferrimagnetic insulators. Magnetic field controls transport through modifications of the magnon dispersion and group velocity, while temperature suppresses transport through exchange stiffness renormalization and enhanced magnon scattering. Thus, the propagation of thermal magnons is determined not simply by their population, but by the microscopic dispersion and relaxation processes that govern their ability to transport spin. These findings identify magnon group velocity and exchange stiffness as key parameters for controlling heat-driven spin currents in magnetic insulators.

In conclusion, we establish a microscopic framework for thermal magnon transport in low-damping LAFO that extends beyond descriptions based solely on thermal magnon population. Nonlocal spin Seebeck measurements show that the magnetic field suppresses thermal magnon transport through a reduction of the magnon spin diffusion length. Direct measurements of the BV magnon dispersion reveal that this behavior originates from a field-induced reduction in magnon group velocity. Temperature-dependent measurements reveal a complementary mechanism: despite an increasing thermal magnon population with increasing temperature, transport decreases due to the temperature dependence of the exchange stiffness and enhanced magnon scattering. These results establish group velocity and exchange interactions as key parameters for engineering heat-driven spin currents in magnetic insulators.

\subsubsection*{}
The experimental work (materials synthesis, structural and magnetic characterization and non-local transport measurements) was primarily supported by the U.S. Department of Energy, Director, Office of Science, Office of Basic Energy Sciences, Division of Materials Sciences and Engineering under Contract \# DESC0008505. Brillouin Light Scattering measurements (J.T., X.L., X.L.), micromagnetic simulations (K.H., M.F.), and theoretical modeling (G.Q., C.Y.) were supported as part of the Center for Energy Efficient Magnonics, an Energy Frontier Research Center (CEEMag) funded by the U.S. Department of Energy, Office of Science, Basic Energy Sciences at SLAC National Laboratory under contract \# DE-AC02-76SF00515. Part of this work was performed at nano@stanford RRID:SCR\_026695. Research on thermal imaging at UCSD was supported as part of the Quantum Materials for Energy Efficient Neuromorphic Computing (Q-MEEN-C), an Energy Frontier Research Center funded by the U.S. Department of Energy, Office of Science, Basic Energy Sciences under Award \# DE-SC0019273.

\subsubsection*{}
The authors have no conflict of interest.

\subsubsection*{}
The data supporting the findings of this article are openly available in the Stanford Digital Repository at \url{https://doi.org/10.25740/fk354mx0324}.

\bibliography{ref}

\begin{thebibliography}{51}%
\makeatletter
\providecommand \@ifxundefined [1]{%
 \@ifx{#1\undefined}
}%
\providecommand \@ifnum [1]{%
 \ifnum #1\expandafter \@firstoftwo
 \else \expandafter \@secondoftwo
 \fi
}%
\providecommand \@ifx [1]{%
 \ifx #1\expandafter \@firstoftwo
 \else \expandafter \@secondoftwo
 \fi
}%
\providecommand \natexlab [1]{#1}%
\providecommand \enquote  [1]{``#1''}%
\providecommand \bibnamefont  [1]{#1}%
\providecommand \bibfnamefont [1]{#1}%
\providecommand \citenamefont [1]{#1}%
\providecommand \href@noop [0]{\@secondoftwo}%
\providecommand \href [0]{\begingroup \@sanitize@url \@href}%
\providecommand \@href[1]{\@@startlink{#1}\@@href}%
\providecommand \@@href[1]{\endgroup#1\@@endlink}%
\providecommand \@sanitize@url [0]{\catcode `\\12\catcode `\$12\catcode `\&12\catcode `\#12\catcode `\^12\catcode `\_12\catcode `\%12\relax}%
\providecommand \@@startlink[1]{}%
\providecommand \@@endlink[0]{}%
\providecommand \url  [0]{\begingroup\@sanitize@url \@url }%
\providecommand \@url [1]{\endgroup\@href {#1}{\urlprefix }}%
\providecommand \urlprefix  [0]{URL }%
\providecommand \Eprint [0]{\href }%
\providecommand \doibase [0]{https://doi.org/}%
\providecommand \selectlanguage [0]{\@gobble}%
\providecommand \bibinfo  [0]{\@secondoftwo}%
\providecommand \bibfield  [0]{\@secondoftwo}%
\providecommand \translation [1]{[#1]}%
\providecommand \BibitemOpen [0]{}%
\providecommand \bibitemStop [0]{}%
\providecommand \bibitemNoStop [0]{.\EOS\space}%
\providecommand \EOS [0]{\spacefactor3000\relax}%
\providecommand \BibitemShut  [1]{\csname bibitem#1\endcsname}%
\let\auto@bib@innerbib\@empty
\bibitem [{\citenamefont {Demokritov}\ and\ \citenamefont {Slavin}(2012)}]{demokritov2012magnonics}%
  \BibitemOpen
  \bibfield  {author} {\bibinfo {author} {\bibfnamefont {S.~O.}\ \bibnamefont {Demokritov}}\ and\ \bibinfo {author} {\bibfnamefont {A.~N.}\ \bibnamefont {Slavin}},\ }\href@noop {} {\emph {\bibinfo {title} {Magnonics: From fundamentals to applications}}},\ Vol.\ \bibinfo {volume} {125}\ (\bibinfo  {publisher} {Springer Science \& Business Media},\ \bibinfo {year} {2012})\BibitemShut {NoStop}%
\bibitem [{\citenamefont {Hoffmann}\ and\ \citenamefont {Bader}(2015)}]{hoffmann2015opportunities}%
  \BibitemOpen
  \bibfield  {author} {\bibinfo {author} {\bibfnamefont {A.}~\bibnamefont {Hoffmann}}\ and\ \bibinfo {author} {\bibfnamefont {S.~D.}\ \bibnamefont {Bader}},\ }\bibfield  {title} {\bibinfo {title} {Opportunities at the frontiers of spintronics},\ }\href {https://doi.org/10.1103/PhysRevApplied.4.047001} {\bibfield  {journal} {\bibinfo  {journal} {Physical Review Applied}\ }\textbf {\bibinfo {volume} {4}},\ \bibinfo {pages} {047001} (\bibinfo {year} {2015})}\BibitemShut {NoStop}%
\bibitem [{\citenamefont {Moradi}\ \emph {et~al.}(2019)\citenamefont {Moradi}, \citenamefont {Farkhani}, \citenamefont {Zeinali}, \citenamefont {Ghanatian}, \citenamefont {Pelloux-Prayer}, \citenamefont {Boehnert}, \citenamefont {Zahedinejad}, \citenamefont {Heidari}, \citenamefont {Nabaei}, \citenamefont {Ferreira}, \citenamefont {{\AA}kerman},\ and\ \citenamefont {Madsen}}]{Moradi2019}%
  \BibitemOpen
  \bibfield  {author} {\bibinfo {author} {\bibfnamefont {F.}~\bibnamefont {Moradi}}, \bibinfo {author} {\bibfnamefont {H.}~\bibnamefont {Farkhani}}, \bibinfo {author} {\bibfnamefont {B.}~\bibnamefont {Zeinali}}, \bibinfo {author} {\bibfnamefont {H.}~\bibnamefont {Ghanatian}}, \bibinfo {author} {\bibfnamefont {J.}~\bibnamefont {Pelloux-Prayer}}, \bibinfo {author} {\bibfnamefont {T.}~\bibnamefont {Boehnert}}, \bibinfo {author} {\bibfnamefont {M.}~\bibnamefont {Zahedinejad}}, \bibinfo {author} {\bibfnamefont {H.}~\bibnamefont {Heidari}}, \bibinfo {author} {\bibfnamefont {V.}~\bibnamefont {Nabaei}}, \bibinfo {author} {\bibfnamefont {R.}~\bibnamefont {Ferreira}}, \bibinfo {author} {\bibfnamefont {J.}~\bibnamefont {{\AA}kerman}},\ and\ \bibinfo {author} {\bibfnamefont {J.~K.}\ \bibnamefont {Madsen}},\ }\bibfield  {title} {\bibinfo {title} {Spin-orbit-torque-based devices, circuits and architectures},\ }\href@noop {} {\bibfield  {journal} {\bibinfo  {journal} {ArXiv}\ }\textbf {\bibinfo {volume} {abs/1912.01347}}
  (\bibinfo {year} {2019})},\ \Eprint {https://arxiv.org/abs/1912.01347} {arXiv:1912.01347} \BibitemShut {NoStop}%
\bibitem [{\citenamefont {Parkin}\ and\ \citenamefont {Yang}(2015)}]{Parkin2015}%
  \BibitemOpen
  \bibfield  {author} {\bibinfo {author} {\bibfnamefont {S.}~\bibnamefont {Parkin}}\ and\ \bibinfo {author} {\bibfnamefont {S.}~\bibnamefont {Yang}},\ }\bibfield  {title} {\bibinfo {title} {Memory on racetrack},\ }\href {https://doi.org/10.1038/nnano.2015.41} {\bibfield  {journal} {\bibinfo  {journal} {Nature Nanotech}\ }\textbf {\bibinfo {volume} {10}},\ \bibinfo {pages} {195} (\bibinfo {year} {2015})}\BibitemShut {NoStop}%
\bibitem [{\citenamefont {Uchida}\ \emph {et~al.}(2010)\citenamefont {Uchida}, \citenamefont {Xiao}, \citenamefont {Adachi}, \citenamefont {Ohe}, \citenamefont {Takahashi}, \citenamefont {Ieda}, \citenamefont {Ota}, \citenamefont {Kajiwara}, \citenamefont {Umezawa}, \citenamefont {Kawai} \emph {et~al.}}]{uchida2010spin}%
  \BibitemOpen
  \bibfield  {author} {\bibinfo {author} {\bibfnamefont {K.-i.}\ \bibnamefont {Uchida}}, \bibinfo {author} {\bibfnamefont {J.}~\bibnamefont {Xiao}}, \bibinfo {author} {\bibfnamefont {H.}~\bibnamefont {Adachi}}, \bibinfo {author} {\bibfnamefont {J.-i.}\ \bibnamefont {Ohe}}, \bibinfo {author} {\bibfnamefont {S.}~\bibnamefont {Takahashi}}, \bibinfo {author} {\bibfnamefont {J.}~\bibnamefont {Ieda}}, \bibinfo {author} {\bibfnamefont {T.}~\bibnamefont {Ota}}, \bibinfo {author} {\bibfnamefont {Y.}~\bibnamefont {Kajiwara}}, \bibinfo {author} {\bibfnamefont {H.}~\bibnamefont {Umezawa}}, \bibinfo {author} {\bibfnamefont {H.}~\bibnamefont {Kawai}}, \emph {et~al.},\ }\bibfield  {title} {\bibinfo {title} {Spin seebeck insulator},\ }\href {https://doi.org/nmat2856} {\bibfield  {journal} {\bibinfo  {journal} {Nature materials}\ }\textbf {\bibinfo {volume} {9}},\ \bibinfo {pages} {894} (\bibinfo {year} {2010})}\BibitemShut {NoStop}%
\bibitem [{\citenamefont {Gomez-Perez}\ \emph {et~al.}(2020)\citenamefont {Gomez-Perez}, \citenamefont {V{\'e}lez}, \citenamefont {Hueso},\ and\ \citenamefont {Casanova}}]{gomez2020differences}%
  \BibitemOpen
  \bibfield  {author} {\bibinfo {author} {\bibfnamefont {J.~M.}\ \bibnamefont {Gomez-Perez}}, \bibinfo {author} {\bibfnamefont {S.}~\bibnamefont {V{\'e}lez}}, \bibinfo {author} {\bibfnamefont {L.~E.}\ \bibnamefont {Hueso}},\ and\ \bibinfo {author} {\bibfnamefont {F.}~\bibnamefont {Casanova}},\ }\bibfield  {title} {\bibinfo {title} {Differences in the magnon diffusion length for electrically and thermally driven magnon currents in {{Y}}$_3${{Fe}}$_5${{O}}$_{12}$},\ }\href {https://doi.org/10.1103/PhysRevB.101.184420} {\bibfield  {journal} {\bibinfo  {journal} {Physical Review B}\ }\textbf {\bibinfo {volume} {101}},\ \bibinfo {pages} {184420} (\bibinfo {year} {2020})}\BibitemShut {NoStop}%
\bibitem [{\citenamefont {Chumak}\ \emph {et~al.}(2015)\citenamefont {Chumak}, \citenamefont {Vasyuchka}, \citenamefont {Serga},\ and\ \citenamefont {Hillebrands}}]{chumak2015magnon}%
  \BibitemOpen
  \bibfield  {author} {\bibinfo {author} {\bibfnamefont {A.~V.}\ \bibnamefont {Chumak}}, \bibinfo {author} {\bibfnamefont {V.~I.}\ \bibnamefont {Vasyuchka}}, \bibinfo {author} {\bibfnamefont {A.~A.}\ \bibnamefont {Serga}},\ and\ \bibinfo {author} {\bibfnamefont {B.}~\bibnamefont {Hillebrands}},\ }\bibfield  {title} {\bibinfo {title} {Magnon spintronics},\ }\href {https://doi.org/nphys3347} {\bibfield  {journal} {\bibinfo  {journal} {Nature physics}\ }\textbf {\bibinfo {volume} {11}},\ \bibinfo {pages} {453} (\bibinfo {year} {2015})}\BibitemShut {NoStop}%
\bibitem [{\citenamefont {Giles}\ \emph {et~al.}(2017)\citenamefont {Giles}, \citenamefont {Yang}, \citenamefont {Jamison}, \citenamefont {Gomez-Perez}, \citenamefont {V{\'e}lez}, \citenamefont {Hueso}, \citenamefont {Casanova},\ and\ \citenamefont {Myers}}]{giles2017thermally}%
  \BibitemOpen
  \bibfield  {author} {\bibinfo {author} {\bibfnamefont {B.~L.}\ \bibnamefont {Giles}}, \bibinfo {author} {\bibfnamefont {Z.}~\bibnamefont {Yang}}, \bibinfo {author} {\bibfnamefont {J.~S.}\ \bibnamefont {Jamison}}, \bibinfo {author} {\bibfnamefont {J.~M.}\ \bibnamefont {Gomez-Perez}}, \bibinfo {author} {\bibfnamefont {S.}~\bibnamefont {V{\'e}lez}}, \bibinfo {author} {\bibfnamefont {L.~E.}\ \bibnamefont {Hueso}}, \bibinfo {author} {\bibfnamefont {F.}~\bibnamefont {Casanova}},\ and\ \bibinfo {author} {\bibfnamefont {R.~C.}\ \bibnamefont {Myers}},\ }\bibfield  {title} {\bibinfo {title} {Thermally driven long-range magnon spin currents in yttrium iron garnet due to intrinsic spin seebeck effect},\ }\href {https://doi.org/10.1103/PhysRevB.96.180412} {\bibfield  {journal} {\bibinfo  {journal} {Physical Review B}\ }\textbf {\bibinfo {volume} {96}},\ \bibinfo {pages} {180412} (\bibinfo {year} {2017})}\BibitemShut {NoStop}%
\bibitem [{\citenamefont {Gao}\ \emph {et~al.}(2022)\citenamefont {Gao}, \citenamefont {Lambert}, \citenamefont {Schlitz}, \citenamefont {Fiebig}, \citenamefont {Gambardella},\ and\ \citenamefont {V{\'e}lez}}]{gao2022magnon}%
  \BibitemOpen
  \bibfield  {author} {\bibinfo {author} {\bibfnamefont {J.}~\bibnamefont {Gao}}, \bibinfo {author} {\bibfnamefont {C.-H.}\ \bibnamefont {Lambert}}, \bibinfo {author} {\bibfnamefont {R.}~\bibnamefont {Schlitz}}, \bibinfo {author} {\bibfnamefont {M.}~\bibnamefont {Fiebig}}, \bibinfo {author} {\bibfnamefont {P.}~\bibnamefont {Gambardella}},\ and\ \bibinfo {author} {\bibfnamefont {S.}~\bibnamefont {V{\'e}lez}},\ }\bibfield  {title} {\bibinfo {title} {Magnon transport and thermoelectric effects in ultrathin {{Tm}}$_3${{Fe}}$_5${{O}}$_{12}$/{{Pt}} nonlocal devices},\ }\href {https://doi.org/10.1103/PhysRevResearch.4.043214} {\bibfield  {journal} {\bibinfo  {journal} {Physical Review Research}\ }\textbf {\bibinfo {volume} {4}},\ \bibinfo {pages} {043214} (\bibinfo {year} {2022})}\BibitemShut {NoStop}%
\bibitem [{\citenamefont {Meyer}\ \emph {et~al.}(2017)\citenamefont {Meyer}, \citenamefont {Chen}, \citenamefont {Wimmer}, \citenamefont {Althammer}, \citenamefont {Wimmer}, \citenamefont {Schlitz}, \citenamefont {Gepr{\"a}gs}, \citenamefont {Huebl}, \citenamefont {K{\"o}dderitzsch}, \citenamefont {Ebert} \emph {et~al.}}]{meyer2017observation}%
  \BibitemOpen
  \bibfield  {author} {\bibinfo {author} {\bibfnamefont {S.}~\bibnamefont {Meyer}}, \bibinfo {author} {\bibfnamefont {Y.-T.}\ \bibnamefont {Chen}}, \bibinfo {author} {\bibfnamefont {S.}~\bibnamefont {Wimmer}}, \bibinfo {author} {\bibfnamefont {M.}~\bibnamefont {Althammer}}, \bibinfo {author} {\bibfnamefont {T.}~\bibnamefont {Wimmer}}, \bibinfo {author} {\bibfnamefont {R.}~\bibnamefont {Schlitz}}, \bibinfo {author} {\bibfnamefont {S.}~\bibnamefont {Gepr{\"a}gs}}, \bibinfo {author} {\bibfnamefont {H.}~\bibnamefont {Huebl}}, \bibinfo {author} {\bibfnamefont {D.}~\bibnamefont {K{\"o}dderitzsch}}, \bibinfo {author} {\bibfnamefont {H.}~\bibnamefont {Ebert}}, \emph {et~al.},\ }\bibfield  {title} {\bibinfo {title} {Observation of the spin nernst effect},\ }\href {https://doi.org/10.1038/nmat4964} {\bibfield  {journal} {\bibinfo  {journal} {Nature materials}\ }\textbf {\bibinfo {volume} {16}},\ \bibinfo {pages} {977} (\bibinfo {year} {2017})}\BibitemShut {NoStop}%
\bibitem [{\citenamefont {Cornelissen}\ and\ \citenamefont {Van~Wees}(2016)}]{cornelissen2016magnetic}%
  \BibitemOpen
  \bibfield  {author} {\bibinfo {author} {\bibfnamefont {L.}~\bibnamefont {Cornelissen}}\ and\ \bibinfo {author} {\bibfnamefont {B.}~\bibnamefont {Van~Wees}},\ }\bibfield  {title} {\bibinfo {title} {Magnetic field dependence of the magnon spin diffusion length in the magnetic insulator yttrium iron garnet},\ }\href {https://doi.org/10.1103/PhysRevB.93.020403} {\bibfield  {journal} {\bibinfo  {journal} {Physical Review B}\ }\textbf {\bibinfo {volume} {93}},\ \bibinfo {pages} {020403} (\bibinfo {year} {2016})}\BibitemShut {NoStop}%
\bibitem [{\citenamefont {Cornelissen}\ \emph {et~al.}(2017)\citenamefont {Cornelissen}, \citenamefont {Oyanagi}, \citenamefont {Kikkawa}, \citenamefont {Qiu}, \citenamefont {Kuschel}, \citenamefont {Bauer}, \citenamefont {Van~Wees},\ and\ \citenamefont {Saitoh}}]{cornelissen2017nonlocal}%
  \BibitemOpen
  \bibfield  {author} {\bibinfo {author} {\bibfnamefont {L.}~\bibnamefont {Cornelissen}}, \bibinfo {author} {\bibfnamefont {K.}~\bibnamefont {Oyanagi}}, \bibinfo {author} {\bibfnamefont {T.}~\bibnamefont {Kikkawa}}, \bibinfo {author} {\bibfnamefont {Z.}~\bibnamefont {Qiu}}, \bibinfo {author} {\bibfnamefont {T.}~\bibnamefont {Kuschel}}, \bibinfo {author} {\bibfnamefont {G.}~\bibnamefont {Bauer}}, \bibinfo {author} {\bibfnamefont {B.}~\bibnamefont {Van~Wees}},\ and\ \bibinfo {author} {\bibfnamefont {E.}~\bibnamefont {Saitoh}},\ }\bibfield  {title} {\bibinfo {title} {Nonlocal magnon-polaron transport in yttrium iron garnet},\ }\href {https://doi.org/10.1103/PhysRevB.96.104441} {\bibfield  {journal} {\bibinfo  {journal} {Physical Review B}\ }\textbf {\bibinfo {volume} {96}},\ \bibinfo {pages} {104441} (\bibinfo {year} {2017})}\BibitemShut {NoStop}%
\bibitem [{\citenamefont {Oyanagi}\ \emph {et~al.}(2020)\citenamefont {Oyanagi}, \citenamefont {Kikkawa},\ and\ \citenamefont {Saitoh}}]{oyanagi2020magnetic}%
  \BibitemOpen
  \bibfield  {author} {\bibinfo {author} {\bibfnamefont {K.}~\bibnamefont {Oyanagi}}, \bibinfo {author} {\bibfnamefont {T.}~\bibnamefont {Kikkawa}},\ and\ \bibinfo {author} {\bibfnamefont {E.}~\bibnamefont {Saitoh}},\ }\bibfield  {title} {\bibinfo {title} {Magnetic field dependence of the nonlocal spin seebeck effect in {{Pt}}/{{YIG}}/{{Pt}} systems at low temperatures},\ }\bibfield  {journal} {\bibinfo  {journal} {AIP Advances}\ }\textbf {\bibinfo {volume} {10}},\ \href {https://doi.org/10.1063/1.5135944} {10.1063/1.5135944} (\bibinfo {year} {2020})\BibitemShut {NoStop}%
\bibitem [{\citenamefont {Sola}\ \emph {et~al.}(2015)\citenamefont {Sola}, \citenamefont {Kuepferling}, \citenamefont {Basso}, \citenamefont {Pasquale}, \citenamefont {Kikkawa}, \citenamefont {Uchida},\ and\ \citenamefont {Saitoh}}]{sola2015evaluation}%
  \BibitemOpen
  \bibfield  {author} {\bibinfo {author} {\bibfnamefont {A.}~\bibnamefont {Sola}}, \bibinfo {author} {\bibfnamefont {M.}~\bibnamefont {Kuepferling}}, \bibinfo {author} {\bibfnamefont {V.}~\bibnamefont {Basso}}, \bibinfo {author} {\bibfnamefont {M.}~\bibnamefont {Pasquale}}, \bibinfo {author} {\bibfnamefont {T.}~\bibnamefont {Kikkawa}}, \bibinfo {author} {\bibfnamefont {K.}~\bibnamefont {Uchida}},\ and\ \bibinfo {author} {\bibfnamefont {E.}~\bibnamefont {Saitoh}},\ }\bibfield  {title} {\bibinfo {title} {Evaluation of thermal gradients in longitudinal spin seebeck effect measurements},\ }\bibfield  {journal} {\bibinfo  {journal} {Journal of Applied Physics}\ }\textbf {\bibinfo {volume} {117}},\ \href {https://doi.org/10.1063/1.4916762} {10.1063/1.4916762} (\bibinfo {year} {2015})\BibitemShut {NoStop}%
\bibitem [{\citenamefont {Vlietstra}\ \emph {et~al.}(2014)\citenamefont {Vlietstra}, \citenamefont {Shan}, \citenamefont {van Wees}, \citenamefont {Isasa}, \citenamefont {Casanova},\ and\ \citenamefont {Ben~Youssef}}]{vlietstra2014simultaneous}%
  \BibitemOpen
  \bibfield  {author} {\bibinfo {author} {\bibfnamefont {N.}~\bibnamefont {Vlietstra}}, \bibinfo {author} {\bibfnamefont {J.}~\bibnamefont {Shan}}, \bibinfo {author} {\bibfnamefont {B.~J.}\ \bibnamefont {van Wees}}, \bibinfo {author} {\bibfnamefont {M.}~\bibnamefont {Isasa}}, \bibinfo {author} {\bibfnamefont {F.}~\bibnamefont {Casanova}},\ and\ \bibinfo {author} {\bibfnamefont {J.}~\bibnamefont {Ben~Youssef}},\ }\bibfield  {title} {\bibinfo {title} {Simultaneous detection of the spin-hall magnetoresistance and the spin-seebeck effect in platinum and tantalum on yttrium iron garnet},\ }\href {https://doi.org/10.1103/PhysRevB.90.174436} {\bibfield  {journal} {\bibinfo  {journal} {Physical Review B}\ }\textbf {\bibinfo {volume} {90}},\ \bibinfo {pages} {174436} (\bibinfo {year} {2014})}\BibitemShut {NoStop}%
\bibitem [{\citenamefont {Cosset-Ch{\'e}neau}\ \emph {et~al.}(2024)\citenamefont {Cosset-Ch{\'e}neau}, \citenamefont {Tirion}, \citenamefont {Wei}, \citenamefont {Ben~Youssef},\ and\ \citenamefont {Van~Wees}}]{cosset2024nonreciprocal}%
  \BibitemOpen
  \bibfield  {author} {\bibinfo {author} {\bibfnamefont {M.}~\bibnamefont {Cosset-Ch{\'e}neau}}, \bibinfo {author} {\bibfnamefont {S.}~\bibnamefont {Tirion}}, \bibinfo {author} {\bibfnamefont {X.-Y.}\ \bibnamefont {Wei}}, \bibinfo {author} {\bibfnamefont {J.}~\bibnamefont {Ben~Youssef}},\ and\ \bibinfo {author} {\bibfnamefont {B.}~\bibnamefont {Van~Wees}},\ }\bibfield  {title} {\bibinfo {title} {Nonreciprocal transport of thermally generated magnons},\ }\href {https://doi.org/10.1103/PhysRevB.110.214418} {\bibfield  {journal} {\bibinfo  {journal} {Physical Review B}\ }\textbf {\bibinfo {volume} {110}},\ \bibinfo {pages} {214418} (\bibinfo {year} {2024})}\BibitemShut {NoStop}%
\bibitem [{\citenamefont {Weiler}\ \emph {et~al.}(2012)\citenamefont {Weiler}, \citenamefont {Althammer}, \citenamefont {Czeschka}, \citenamefont {Huebl}, \citenamefont {Wagner}, \citenamefont {Opel}, \citenamefont {Imort}, \citenamefont {Reiss}, \citenamefont {Thomas}, \citenamefont {Gross} \emph {et~al.}}]{weiler2012local}%
  \BibitemOpen
  \bibfield  {author} {\bibinfo {author} {\bibfnamefont {M.}~\bibnamefont {Weiler}}, \bibinfo {author} {\bibfnamefont {M.}~\bibnamefont {Althammer}}, \bibinfo {author} {\bibfnamefont {F.~D.}\ \bibnamefont {Czeschka}}, \bibinfo {author} {\bibfnamefont {H.}~\bibnamefont {Huebl}}, \bibinfo {author} {\bibfnamefont {M.~S.}\ \bibnamefont {Wagner}}, \bibinfo {author} {\bibfnamefont {M.}~\bibnamefont {Opel}}, \bibinfo {author} {\bibfnamefont {I.-M.}\ \bibnamefont {Imort}}, \bibinfo {author} {\bibfnamefont {G.}~\bibnamefont {Reiss}}, \bibinfo {author} {\bibfnamefont {A.}~\bibnamefont {Thomas}}, \bibinfo {author} {\bibfnamefont {R.}~\bibnamefont {Gross}}, \emph {et~al.},\ }\bibfield  {title} {\bibinfo {title} {Local charge and spin currents in magnetothermal landscapes},\ }\href {https://doi.org/10.1103/PhysRevLett.108.106602} {\bibfield  {journal} {\bibinfo  {journal} {Physical review letters}\ }\textbf {\bibinfo {volume} {108}},\ \bibinfo {pages} {106602} (\bibinfo {year} {2012})}\BibitemShut {NoStop}%
\bibitem [{\citenamefont {Taghinejad}\ \emph {et~al.}(2025)\citenamefont {Taghinejad}, \citenamefont {Yamakawa}, \citenamefont {Huang}, \citenamefont {Lyu}, \citenamefont {Pritchard~Cairns}, \citenamefont {Husain}, \citenamefont {Ramesh},\ and\ \citenamefont {Analytis}}]{taghinejad2025low}%
  \BibitemOpen
  \bibfield  {author} {\bibinfo {author} {\bibfnamefont {H.}~\bibnamefont {Taghinejad}}, \bibinfo {author} {\bibfnamefont {K.}~\bibnamefont {Yamakawa}}, \bibinfo {author} {\bibfnamefont {X.}~\bibnamefont {Huang}}, \bibinfo {author} {\bibfnamefont {Y.}~\bibnamefont {Lyu}}, \bibinfo {author} {\bibfnamefont {L.}~\bibnamefont {Pritchard~Cairns}}, \bibinfo {author} {\bibfnamefont {S.}~\bibnamefont {Husain}}, \bibinfo {author} {\bibfnamefont {R.}~\bibnamefont {Ramesh}},\ and\ \bibinfo {author} {\bibfnamefont {J.~G.}\ \bibnamefont {Analytis}},\ }\bibfield  {title} {\bibinfo {title} {Low-field regime of magnon transport in pld-grown yig films},\ }\href {https://doi.org/10.1021/acs.nanolett.4c06592} {\bibfield  {journal} {\bibinfo  {journal} {Nano Letters}\ }\textbf {\bibinfo {volume} {25}},\ \bibinfo {pages} {6438} (\bibinfo {year} {2025})}\BibitemShut {NoStop}%
\bibitem [{\citenamefont {Kurfman}\ \emph {et~al.}(2026)\citenamefont {Kurfman}, \citenamefont {Candido}, \citenamefont {Wooten}, \citenamefont {Zheng}, \citenamefont {Newburger}, \citenamefont {Cheng}, \citenamefont {Kawakami}, \citenamefont {Heremans}, \citenamefont {Flatt{\'e}},\ and\ \citenamefont {Johnston-Halperin}}]{kurfman2026magnon}%
  \BibitemOpen
  \bibfield  {author} {\bibinfo {author} {\bibfnamefont {S.~W.}\ \bibnamefont {Kurfman}}, \bibinfo {author} {\bibfnamefont {D.~R.}\ \bibnamefont {Candido}}, \bibinfo {author} {\bibfnamefont {B.}~\bibnamefont {Wooten}}, \bibinfo {author} {\bibfnamefont {Y.}~\bibnamefont {Zheng}}, \bibinfo {author} {\bibfnamefont {M.~J.}\ \bibnamefont {Newburger}}, \bibinfo {author} {\bibfnamefont {S.}~\bibnamefont {Cheng}}, \bibinfo {author} {\bibfnamefont {R.~K.}\ \bibnamefont {Kawakami}}, \bibinfo {author} {\bibfnamefont {J.~P.}\ \bibnamefont {Heremans}}, \bibinfo {author} {\bibfnamefont {M.~E.}\ \bibnamefont {Flatt{\'e}}},\ and\ \bibinfo {author} {\bibfnamefont {E.}~\bibnamefont {Johnston-Halperin}},\ }\bibfield  {title} {\bibinfo {title} {Magnon diffusion length and longitudinal spin seebeck effect in vanadium tetracyanoethylene v [tcne] x (x~ 2)},\ }\href@noop {} {\bibfield  {journal} {\bibinfo  {journal} {Physical Review B}\ }\textbf {\bibinfo {volume} {113}},\ \bibinfo {pages} {L220408} (\bibinfo {year}
  {2026})}\BibitemShut {NoStop}%
\bibitem [{\citenamefont {Qin}\ \emph {et~al.}(2018)\citenamefont {Qin}, \citenamefont {H{\"a}m{\"a}l{\"a}inen}, \citenamefont {Arjas}, \citenamefont {Witteveen},\ and\ \citenamefont {Van~Dijken}}]{qin2018propagating}%
  \BibitemOpen
  \bibfield  {author} {\bibinfo {author} {\bibfnamefont {H.}~\bibnamefont {Qin}}, \bibinfo {author} {\bibfnamefont {S.~J.}\ \bibnamefont {H{\"a}m{\"a}l{\"a}inen}}, \bibinfo {author} {\bibfnamefont {K.}~\bibnamefont {Arjas}}, \bibinfo {author} {\bibfnamefont {J.}~\bibnamefont {Witteveen}},\ and\ \bibinfo {author} {\bibfnamefont {S.}~\bibnamefont {Van~Dijken}},\ }\bibfield  {title} {\bibinfo {title} {Propagating spin waves in nanometer-thick yttrium iron garnet films: Dependence on wave vector, magnetic field strength, and angle},\ }\href {https://doi.org/10.1103/PhysRevB.98.224422} {\bibfield  {journal} {\bibinfo  {journal} {Physical Review B}\ }\textbf {\bibinfo {volume} {98}},\ \bibinfo {pages} {224422} (\bibinfo {year} {2018})}\BibitemShut {NoStop}%
\bibitem [{\citenamefont {Krysztofik}\ \emph {et~al.}(2017)\citenamefont {Krysztofik}, \citenamefont {G{\l}owi{\'n}ski}, \citenamefont {Ku{\'s}wik}, \citenamefont {Zi{\k{e}}tek}, \citenamefont {Coy}, \citenamefont {Rych{\l}y}, \citenamefont {Jurga}, \citenamefont {Stobiecki},\ and\ \citenamefont {Dubowik}}]{krysztofik2017characterization}%
  \BibitemOpen
  \bibfield  {author} {\bibinfo {author} {\bibfnamefont {A.}~\bibnamefont {Krysztofik}}, \bibinfo {author} {\bibfnamefont {H.}~\bibnamefont {G{\l}owi{\'n}ski}}, \bibinfo {author} {\bibfnamefont {P.}~\bibnamefont {Ku{\'s}wik}}, \bibinfo {author} {\bibfnamefont {S.}~\bibnamefont {Zi{\k{e}}tek}}, \bibinfo {author} {\bibfnamefont {L.~E.}\ \bibnamefont {Coy}}, \bibinfo {author} {\bibfnamefont {J.~N.}\ \bibnamefont {Rych{\l}y}}, \bibinfo {author} {\bibfnamefont {S.}~\bibnamefont {Jurga}}, \bibinfo {author} {\bibfnamefont {T.~W.}\ \bibnamefont {Stobiecki}},\ and\ \bibinfo {author} {\bibfnamefont {J.}~\bibnamefont {Dubowik}},\ }\bibfield  {title} {\bibinfo {title} {Characterization of spin wave propagation in (1 1 1) yig thin films with large anisotropy},\ }\href {https://doi.org/10.1088/1361-6463/aa6df0} {\bibfield  {journal} {\bibinfo  {journal} {Journal of Physics D: Applied Physics}\ }\textbf {\bibinfo {volume} {50}},\ \bibinfo {pages} {235004} (\bibinfo {year} {2017})}\BibitemShut {NoStop}%
\bibitem [{\citenamefont {Shiota}\ \emph {et~al.}(2026)\citenamefont {Shiota}, \citenamefont {Kan}, \citenamefont {Hisatomi}, \citenamefont {Karube}, \citenamefont {Shimakawa},\ and\ \citenamefont {Ono}}]{shiota2026perpendicular}%
  \BibitemOpen
  \bibfield  {author} {\bibinfo {author} {\bibfnamefont {Y.}~\bibnamefont {Shiota}}, \bibinfo {author} {\bibfnamefont {D.}~\bibnamefont {Kan}}, \bibinfo {author} {\bibfnamefont {R.}~\bibnamefont {Hisatomi}}, \bibinfo {author} {\bibfnamefont {S.}~\bibnamefont {Karube}}, \bibinfo {author} {\bibfnamefont {Y.}~\bibnamefont {Shimakawa}},\ and\ \bibinfo {author} {\bibfnamefont {T.}~\bibnamefont {Ono}},\ }\bibfield  {title} {\bibinfo {title} {Perpendicular magnetization in sputtered yttrium iron garnet thin films for spin-wave propagation and spin-orbit-torque switching},\ }\href {https://doi.org/10.1103/32hy-k2mn} {\bibfield  {journal} {\bibinfo  {journal} {Physical Review Applied}\ }\textbf {\bibinfo {volume} {25}},\ \bibinfo {pages} {044077} (\bibinfo {year} {2026})}\BibitemShut {NoStop}%
\bibitem [{sup(2027)}]{supp}%
  \BibitemOpen
  \href@noop {} {\bibinfo {title} {Supplemntal information}} (\bibinfo {year} {2027}),\ \bibinfo {note} {see Supplemental Material at [URL will be inserted by publisher] Sample growth and device fabrication, structural and magnetic characterization, thermal imaging measurements, extraction of the spin diffusion lenght, the dispersion relation at different magnetic fields and micromagneti simulation parameters and method}\BibitemShut {NoStop}%
\bibitem [{\citenamefont {O'Mahoney}\ \emph {et~al.}(2023)\citenamefont {O'Mahoney}, \citenamefont {Channa}, \citenamefont {Zheng}, \citenamefont {Vailionis}, \citenamefont {Shafer}, \citenamefont {Klewe}, \citenamefont {Suzuki} \emph {et~al.}}]{daisy2023aluminum}%
  \BibitemOpen
  \bibfield  {author} {\bibinfo {author} {\bibfnamefont {D.}~\bibnamefont {O'Mahoney}}, \bibinfo {author} {\bibfnamefont {S.}~\bibnamefont {Channa}}, \bibinfo {author} {\bibfnamefont {X.~Y.}\ \bibnamefont {Zheng}}, \bibinfo {author} {\bibfnamefont {A.}~\bibnamefont {Vailionis}}, \bibinfo {author} {\bibfnamefont {P.}~\bibnamefont {Shafer}}, \bibinfo {author} {\bibfnamefont {C.}~\bibnamefont {Klewe}}, \bibinfo {author} {\bibfnamefont {Y.}~\bibnamefont {Suzuki}}, \emph {et~al.},\ }\bibfield  {title} {\bibinfo {title} {Aluminum substitution in low damping epitaxial lithium ferrite films},\ }\bibfield  {journal} {\bibinfo  {journal} {Applied Physics Letters}\ }\textbf {\bibinfo {volume} {123}},\ \href {https://doi.org/10.1063/5.0163362} {10.1063/5.0163362} (\bibinfo {year} {2023})\BibitemShut {NoStop}%
\bibitem [{\citenamefont {Zheng}\ \emph {et~al.}(2023)\citenamefont {Zheng}, \citenamefont {Channa}, \citenamefont {Riddiford}, \citenamefont {Wisser}, \citenamefont {Mahalingam}, \citenamefont {Bowers}, \citenamefont {McConney}, \citenamefont {N’Diaye}, \citenamefont {Vailionis}, \citenamefont {Cogulu} \emph {et~al.}}]{zheng2023ultra}%
  \BibitemOpen
  \bibfield  {author} {\bibinfo {author} {\bibfnamefont {X.~Y.}\ \bibnamefont {Zheng}}, \bibinfo {author} {\bibfnamefont {S.}~\bibnamefont {Channa}}, \bibinfo {author} {\bibfnamefont {L.~J.}\ \bibnamefont {Riddiford}}, \bibinfo {author} {\bibfnamefont {J.~J.}\ \bibnamefont {Wisser}}, \bibinfo {author} {\bibfnamefont {K.}~\bibnamefont {Mahalingam}}, \bibinfo {author} {\bibfnamefont {C.~T.}\ \bibnamefont {Bowers}}, \bibinfo {author} {\bibfnamefont {M.~E.}\ \bibnamefont {McConney}}, \bibinfo {author} {\bibfnamefont {A.~T.}\ \bibnamefont {N’Diaye}}, \bibinfo {author} {\bibfnamefont {A.}~\bibnamefont {Vailionis}}, \bibinfo {author} {\bibfnamefont {E.}~\bibnamefont {Cogulu}}, \emph {et~al.},\ }\bibfield  {title} {\bibinfo {title} {Ultra-thin lithium aluminate spinel ferrite films with perpendicular magnetic anisotropy and low damping},\ }\href {https://doi.org/s41467-023-40733-9} {\bibfield  {journal} {\bibinfo  {journal} {Nature communications}\ }\textbf {\bibinfo {volume} {14}},\ \bibinfo {pages} {4918}
  (\bibinfo {year} {2023})}\BibitemShut {NoStop}%
\bibitem [{\citenamefont {Takana}\ \emph {et~al.}(2025)\citenamefont {Takana}, \citenamefont {Channa}, \citenamefont {Zheng}, \citenamefont {O'Mahoney}, \citenamefont {Alaei}, \citenamefont {Li}, \citenamefont {Vailionis}, \citenamefont {Shafer}, \citenamefont {Klewe}, \citenamefont {Fisher} \emph {et~al.}}]{takana2025low}%
  \BibitemOpen
  \bibfield  {author} {\bibinfo {author} {\bibfnamefont {L.}~\bibnamefont {Takana}}, \bibinfo {author} {\bibfnamefont {S.}~\bibnamefont {Channa}}, \bibinfo {author} {\bibfnamefont {X.~Y.}\ \bibnamefont {Zheng}}, \bibinfo {author} {\bibfnamefont {D.}~\bibnamefont {O'Mahoney}}, \bibinfo {author} {\bibfnamefont {S.}~\bibnamefont {Alaei}}, \bibinfo {author} {\bibfnamefont {Y.}~\bibnamefont {Li}}, \bibinfo {author} {\bibfnamefont {A.}~\bibnamefont {Vailionis}}, \bibinfo {author} {\bibfnamefont {P.}~\bibnamefont {Shafer}}, \bibinfo {author} {\bibfnamefont {C.}~\bibnamefont {Klewe}}, \bibinfo {author} {\bibfnamefont {I.}~\bibnamefont {Fisher}}, \emph {et~al.},\ }\bibfield  {title} {\bibinfo {title} {Low damping (111) oriented lithium aluminum ferrite thin films for spin wave applications},\ }\bibfield  {journal} {\bibinfo  {journal} {Applied Physics Letters}\ }\textbf {\bibinfo {volume} {127}},\ \href {https://doi.org/10.1063/5.0278599} {10.1063/5.0278599} (\bibinfo {year} {2025})\BibitemShut {NoStop}%
\bibitem [{\citenamefont {Zheng}\ \emph {et~al.}(2020)\citenamefont {Zheng}, \citenamefont {Riddiford}, \citenamefont {Wisser}, \citenamefont {Emori},\ and\ \citenamefont {Suzuki}}]{zheng2020ultra}%
  \BibitemOpen
  \bibfield  {author} {\bibinfo {author} {\bibfnamefont {X.~Y.}\ \bibnamefont {Zheng}}, \bibinfo {author} {\bibfnamefont {L.~J.}\ \bibnamefont {Riddiford}}, \bibinfo {author} {\bibfnamefont {J.~J.}\ \bibnamefont {Wisser}}, \bibinfo {author} {\bibfnamefont {S.}~\bibnamefont {Emori}},\ and\ \bibinfo {author} {\bibfnamefont {Y.}~\bibnamefont {Suzuki}},\ }\bibfield  {title} {\bibinfo {title} {Ultra-low magnetic damping in epitaxial {{Li}}$_{0.5}${{Fe}}$_{2.5}${{O}}$_4$ thin films},\ }\bibfield  {journal} {\bibinfo  {journal} {Applied Physics Letters}\ }\textbf {\bibinfo {volume} {117}},\ \href {https://doi.org/10.1063/5.0023077} {10.1063/5.0023077} (\bibinfo {year} {2020})\BibitemShut {NoStop}%
\bibitem [{\citenamefont {Hofer}\ \emph {et~al.}(2025)\citenamefont {Hofer}, \citenamefont {Basaran}, \citenamefont {Wang}, \citenamefont {Torres},\ and\ \citenamefont {Schuller}}]{hofer2025mechanically}%
  \BibitemOpen
  \bibfield  {author} {\bibinfo {author} {\bibfnamefont {J.~A.}\ \bibnamefont {Hofer}}, \bibinfo {author} {\bibfnamefont {A.~C.}\ \bibnamefont {Basaran}}, \bibinfo {author} {\bibfnamefont {T.~D.}\ \bibnamefont {Wang}}, \bibinfo {author} {\bibfnamefont {F.~E.}\ \bibnamefont {Torres}},\ and\ \bibinfo {author} {\bibfnamefont {I.~K.}\ \bibnamefont {Schuller}},\ }\bibfield  {title} {\bibinfo {title} {Mechanically tunable metal--insulator transition in flexible {{VO}}$_2$ devices for ultra-low power electronics},\ }\href {https://doi.org/10.1021/acsaelm.5c01721} {\bibfield  {journal} {\bibinfo  {journal} {ACS Applied Electronic Materials}\ }\textbf {\bibinfo {volume} {7}},\ \bibinfo {pages} {9869} (\bibinfo {year} {2025})}\BibitemShut {NoStop}%
\bibitem [{\citenamefont {Maekawa}\ \emph {et~al.}(2013)\citenamefont {Maekawa}, \citenamefont {Adachi}, \citenamefont {Uchida}, \citenamefont {Ieda},\ and\ \citenamefont {Saitoh}}]{maekawa2013spin}%
  \BibitemOpen
  \bibfield  {author} {\bibinfo {author} {\bibfnamefont {S.}~\bibnamefont {Maekawa}}, \bibinfo {author} {\bibfnamefont {H.}~\bibnamefont {Adachi}}, \bibinfo {author} {\bibfnamefont {K.-i.}\ \bibnamefont {Uchida}}, \bibinfo {author} {\bibfnamefont {J.}~\bibnamefont {Ieda}},\ and\ \bibinfo {author} {\bibfnamefont {E.}~\bibnamefont {Saitoh}},\ }\bibfield  {title} {\bibinfo {title} {Spin current: Experimental and theoretical aspects},\ }\href {https://doi.org/10.7566/JPSJ.82.102002} {\bibfield  {journal} {\bibinfo  {journal} {Journal of the Physical Society of Japan}\ }\textbf {\bibinfo {volume} {82}},\ \bibinfo {pages} {102002} (\bibinfo {year} {2013})}\BibitemShut {NoStop}%
\bibitem [{\citenamefont {Hoffmann}(2007)}]{hoffmann2007pure}%
  \BibitemOpen
  \bibfield  {author} {\bibinfo {author} {\bibfnamefont {A.}~\bibnamefont {Hoffmann}},\ }\bibfield  {title} {\bibinfo {title} {Pure spin-currents},\ }\href {https://doi.org/10.1002/pssc.200775942} {\bibfield  {journal} {\bibinfo  {journal} {physica status solidi c}\ }\textbf {\bibinfo {volume} {4}},\ \bibinfo {pages} {4236} (\bibinfo {year} {2007})}\BibitemShut {NoStop}%
\bibitem [{\citenamefont {Thiery}\ \emph {et~al.}(2017)\citenamefont {Thiery}, \citenamefont {Draveny}, \citenamefont {Naletov}, \citenamefont {Vila}, \citenamefont {Attan{\'e}}, \citenamefont {De~Loubens}, \citenamefont {Viret}, \citenamefont {Beaulieu}, \citenamefont {Youssef}, \citenamefont {Demidov} \emph {et~al.}}]{thiery2017spin}%
  \BibitemOpen
  \bibfield  {author} {\bibinfo {author} {\bibfnamefont {N.}~\bibnamefont {Thiery}}, \bibinfo {author} {\bibfnamefont {A.}~\bibnamefont {Draveny}}, \bibinfo {author} {\bibfnamefont {V.~V.}\ \bibnamefont {Naletov}}, \bibinfo {author} {\bibfnamefont {L.}~\bibnamefont {Vila}}, \bibinfo {author} {\bibfnamefont {J.-P.}\ \bibnamefont {Attan{\'e}}}, \bibinfo {author} {\bibfnamefont {G.}~\bibnamefont {De~Loubens}}, \bibinfo {author} {\bibfnamefont {M.}~\bibnamefont {Viret}}, \bibinfo {author} {\bibfnamefont {N.}~\bibnamefont {Beaulieu}}, \bibinfo {author} {\bibfnamefont {J.~B.}\ \bibnamefont {Youssef}}, \bibinfo {author} {\bibfnamefont {V.~E.}\ \bibnamefont {Demidov}}, \emph {et~al.},\ }\bibfield  {title} {\bibinfo {title} {Spin conductance of {{YIG}} thin films driven from thermal to subthermal magnons regime by large spin-orbit torque},\ }\bibfield  {journal} {\bibinfo  {journal} {arXiv preprint arXiv:1702.05226}\ }\href {https://doi.org/10.48550/arXiv.1702.05226} {10.48550/arXiv.1702.05226} (\bibinfo {year}
  {2017})\BibitemShut {NoStop}%
\bibitem [{\citenamefont {Lebrun}\ \emph {et~al.}(2018)\citenamefont {Lebrun}, \citenamefont {Ross}, \citenamefont {Bender}, \citenamefont {Qaiumzadeh}, \citenamefont {Baldrati}, \citenamefont {Cramer}, \citenamefont {Brataas}, \citenamefont {Duine},\ and\ \citenamefont {Kl{\"a}ui}}]{lebrun2018tunable}%
  \BibitemOpen
  \bibfield  {author} {\bibinfo {author} {\bibfnamefont {R.}~\bibnamefont {Lebrun}}, \bibinfo {author} {\bibfnamefont {A.}~\bibnamefont {Ross}}, \bibinfo {author} {\bibfnamefont {S.}~\bibnamefont {Bender}}, \bibinfo {author} {\bibfnamefont {A.}~\bibnamefont {Qaiumzadeh}}, \bibinfo {author} {\bibfnamefont {L.}~\bibnamefont {Baldrati}}, \bibinfo {author} {\bibfnamefont {J.}~\bibnamefont {Cramer}}, \bibinfo {author} {\bibfnamefont {A.}~\bibnamefont {Brataas}}, \bibinfo {author} {\bibfnamefont {R.}~\bibnamefont {Duine}},\ and\ \bibinfo {author} {\bibfnamefont {M.}~\bibnamefont {Kl{\"a}ui}},\ }\bibfield  {title} {\bibinfo {title} {Tunable long-distance spin transport in a crystalline antiferromagnetic iron oxide},\ }\href {https://doi.org/s41586-018-0490-7} {\bibfield  {journal} {\bibinfo  {journal} {Nature}\ }\textbf {\bibinfo {volume} {561}},\ \bibinfo {pages} {222} (\bibinfo {year} {2018})}\BibitemShut {NoStop}%
\bibitem [{\citenamefont {Kalinikos}\ and\ \citenamefont {Slavin}(1986)}]{kalinikos1986theory}%
  \BibitemOpen
  \bibfield  {author} {\bibinfo {author} {\bibfnamefont {B.~A.}\ \bibnamefont {Kalinikos}}\ and\ \bibinfo {author} {\bibfnamefont {A.~N.}\ \bibnamefont {Slavin}},\ }\bibfield  {title} {\bibinfo {title} {Theory of dipole-exchange spin wave spectrum for ferromagnetic films with mixed exchange boundary conditions},\ }\href@noop {} {\bibfield  {journal} {\bibinfo  {journal} {Journal of Physics C: Solid State Physics}\ }\textbf {\bibinfo {volume} {19}},\ \bibinfo {pages} {7013} (\bibinfo {year} {1986})}\BibitemShut {NoStop}%
\bibitem [{\citenamefont {Tong}\ \emph {et~al.}(2026)\citenamefont {Tong}, \citenamefont {Paudyal}, \citenamefont {Liu}, \citenamefont {Takana}, \citenamefont {Mikhailova}, \citenamefont {Suzuki}, \citenamefont {Paudyal},\ and\ \citenamefont {Li}}]{tong2026direct}%
  \BibitemOpen
  \bibfield  {author} {\bibinfo {author} {\bibfnamefont {J.}~\bibnamefont {Tong}}, \bibinfo {author} {\bibfnamefont {H.}~\bibnamefont {Paudyal}}, \bibinfo {author} {\bibfnamefont {X.}~\bibnamefont {Liu}}, \bibinfo {author} {\bibfnamefont {L.}~\bibnamefont {Takana}}, \bibinfo {author} {\bibfnamefont {K.}~\bibnamefont {Mikhailova}}, \bibinfo {author} {\bibfnamefont {Y.}~\bibnamefont {Suzuki}}, \bibinfo {author} {\bibfnamefont {D.}~\bibnamefont {Paudyal}},\ and\ \bibinfo {author} {\bibfnamefont {X.}~\bibnamefont {Li}},\ }\bibfield  {title} {\bibinfo {title} {Direct observation of tunable magnons in epitaxial lithium aluminum ferrite thin films},\ }\href {https://doi.org/10.1021/acs.nanolett.5c05922} {\bibfield  {journal} {\bibinfo  {journal} {Nano Letters}\ }\textbf {\bibinfo {volume} {26}},\ \bibinfo {pages} {3073} (\bibinfo {year} {2026})}\BibitemShut {NoStop}%
\bibitem [{\citenamefont {Prabhakar}\ and\ \citenamefont {Stancil}(2009)}]{prabhakar2009spin}%
  \BibitemOpen
  \bibfield  {author} {\bibinfo {author} {\bibfnamefont {A.}~\bibnamefont {Prabhakar}}\ and\ \bibinfo {author} {\bibfnamefont {D.~D.}\ \bibnamefont {Stancil}},\ }\href@noop {} {\emph {\bibinfo {title} {Spin waves: Theory and applications}}},\ Vol.~\bibinfo {volume} {5}\ (\bibinfo  {publisher} {Springer},\ \bibinfo {year} {2009})\BibitemShut {NoStop}%
\bibitem [{\citenamefont {Adachi}\ \emph {et~al.}(2010)\citenamefont {Adachi}, \citenamefont {Uchida}, \citenamefont {Saitoh}, \citenamefont {Ohe}, \citenamefont {Takahashi},\ and\ \citenamefont {Maekawa}}]{adachi2010gigantic}%
  \BibitemOpen
  \bibfield  {author} {\bibinfo {author} {\bibfnamefont {H.}~\bibnamefont {Adachi}}, \bibinfo {author} {\bibfnamefont {K.-i.}\ \bibnamefont {Uchida}}, \bibinfo {author} {\bibfnamefont {E.}~\bibnamefont {Saitoh}}, \bibinfo {author} {\bibfnamefont {J.-i.}\ \bibnamefont {Ohe}}, \bibinfo {author} {\bibfnamefont {S.}~\bibnamefont {Takahashi}},\ and\ \bibinfo {author} {\bibfnamefont {S.}~\bibnamefont {Maekawa}},\ }\bibfield  {title} {\bibinfo {title} {Gigantic enhancement of spin seebeck effect by phonon drag},\ }\href@noop {} {\bibfield  {journal} {\bibinfo  {journal} {Applied Physics Letters}\ }\textbf {\bibinfo {volume} {97}} (\bibinfo {year} {2010})}\BibitemShut {NoStop}%
\bibitem [{\citenamefont {Rodr{\'\i}guez-Su{\'a}rez}\ and\ \citenamefont {Rezende}(2023)}]{rodriguez2023dominance}%
  \BibitemOpen
  \bibfield  {author} {\bibinfo {author} {\bibfnamefont {R.~L.}\ \bibnamefont {Rodr{\'\i}guez-Su{\'a}rez}}\ and\ \bibinfo {author} {\bibfnamefont {S.~M.}\ \bibnamefont {Rezende}},\ }\bibfield  {title} {\bibinfo {title} {Dominance of the phonon drag mechanism in the spin seebeck effect at low temperatures},\ }\href {https://doi.org/10.1103/PhysRevB.108.134407} {\bibfield  {journal} {\bibinfo  {journal} {Physical Review B}\ }\textbf {\bibinfo {volume} {108}},\ \bibinfo {pages} {134407} (\bibinfo {year} {2023})}\BibitemShut {NoStop}%
\bibitem [{\citenamefont {Guo}\ \emph {et~al.}(2016)\citenamefont {Guo}, \citenamefont {Cramer}, \citenamefont {Kehlberger}, \citenamefont {Ferguson}, \citenamefont {MacLaren}, \citenamefont {Jakob},\ and\ \citenamefont {Kl{\"a}ui}}]{guo2016influence}%
  \BibitemOpen
  \bibfield  {author} {\bibinfo {author} {\bibfnamefont {E.-J.}\ \bibnamefont {Guo}}, \bibinfo {author} {\bibfnamefont {J.}~\bibnamefont {Cramer}}, \bibinfo {author} {\bibfnamefont {A.}~\bibnamefont {Kehlberger}}, \bibinfo {author} {\bibfnamefont {C.~A.}\ \bibnamefont {Ferguson}}, \bibinfo {author} {\bibfnamefont {D.~A.}\ \bibnamefont {MacLaren}}, \bibinfo {author} {\bibfnamefont {G.}~\bibnamefont {Jakob}},\ and\ \bibinfo {author} {\bibfnamefont {M.}~\bibnamefont {Kl{\"a}ui}},\ }\bibfield  {title} {\bibinfo {title} {Influence of thickness and interface on the low-temperature enhancement of the spin seebeck effect in yig films},\ }\href@noop {} {\bibfield  {journal} {\bibinfo  {journal} {Physical Review X}\ }\textbf {\bibinfo {volume} {6}},\ \bibinfo {pages} {031012} (\bibinfo {year} {2016})}\BibitemShut {NoStop}%
\bibitem [{\citenamefont {vansteenkiste}\ \emph {et~al.}(2014)\citenamefont {vansteenkiste}, \citenamefont {Leliaert}, \citenamefont {Dvornik}, \citenamefont {Helsen}, \citenamefont {Garcia-Sanchez},\ and\ \citenamefont {Waeyenberge}}]{Vansteenkiste2014MuMax}%
  \BibitemOpen
  \bibfield  {author} {\bibinfo {author} {\bibfnamefont {A.}~\bibnamefont {vansteenkiste}}, \bibinfo {author} {\bibfnamefont {J.}~\bibnamefont {Leliaert}}, \bibinfo {author} {\bibfnamefont {M.}~\bibnamefont {Dvornik}}, \bibinfo {author} {\bibfnamefont {M.}~\bibnamefont {Helsen}}, \bibinfo {author} {\bibfnamefont {F.}~\bibnamefont {Garcia-Sanchez}},\ and\ \bibinfo {author} {\bibfnamefont {B.~V.}\ \bibnamefont {Waeyenberge}},\ }\bibfield  {title} {\bibinfo {title} {The design and verification of mumax3},\ }\href {https://doi.org/10.1063/1.4899186} {\bibfield  {journal} {\bibinfo  {journal} {AIP Adv.}\ }\textbf {\bibinfo {volume} {4}},\ \bibinfo {pages} {107133} (\bibinfo {year} {2014})}\BibitemShut {NoStop}%
\bibitem [{\citenamefont {Atxitia}\ \emph {et~al.}(2010)\citenamefont {Atxitia}, \citenamefont {Hinzke}, \citenamefont {Chubykalo-Fesenko}, \citenamefont {Nowak}, \citenamefont {Kachkachi}, \citenamefont {Mryasov}, \citenamefont {Evans},\ and\ \citenamefont {Chantrell}}]{atxitia2010multiscale}%
  \BibitemOpen
  \bibfield  {author} {\bibinfo {author} {\bibfnamefont {U.}~\bibnamefont {Atxitia}}, \bibinfo {author} {\bibfnamefont {D.}~\bibnamefont {Hinzke}}, \bibinfo {author} {\bibfnamefont {O.}~\bibnamefont {Chubykalo-Fesenko}}, \bibinfo {author} {\bibfnamefont {U.}~\bibnamefont {Nowak}}, \bibinfo {author} {\bibfnamefont {H.}~\bibnamefont {Kachkachi}}, \bibinfo {author} {\bibfnamefont {O.~N.}\ \bibnamefont {Mryasov}}, \bibinfo {author} {\bibfnamefont {R.}~\bibnamefont {Evans}},\ and\ \bibinfo {author} {\bibfnamefont {R.~W.}\ \bibnamefont {Chantrell}},\ }\bibfield  {title} {\bibinfo {title} {Multiscale modeling of magnetic materials: Temperature dependence of the exchange stiffness},\ }\href {https://doi.org/10.1103/PhysRevB.82.134440} {\bibfield  {journal} {\bibinfo  {journal} {Physical Review B—Condensed Matter and Materials Physics}\ }\textbf {\bibinfo {volume} {82}},\ \bibinfo {pages} {134440} (\bibinfo {year} {2010})}\BibitemShut {NoStop}%
\bibitem [{\citenamefont {Niitsu}\ \emph {et~al.}(2020)\citenamefont {Niitsu}, \citenamefont {Xu}, \citenamefont {Umetsu}, \citenamefont {Kainuma},\ and\ \citenamefont {Harada}}]{Niitsu_2020}%
  \BibitemOpen
  \bibfield  {author} {\bibinfo {author} {\bibfnamefont {K.}~\bibnamefont {Niitsu}}, \bibinfo {author} {\bibfnamefont {X.}~\bibnamefont {Xu}}, \bibinfo {author} {\bibfnamefont {R.~Y.}\ \bibnamefont {Umetsu}}, \bibinfo {author} {\bibfnamefont {R.}~\bibnamefont {Kainuma}},\ and\ \bibinfo {author} {\bibfnamefont {K.}~\bibnamefont {Harada}},\ }\bibfield  {title} {\bibinfo {title} {Temperature dependence of exchange stiffness in an off-stoichiometric $\mathrm{N}{\mathrm{i}}_{2}\mathrm{MnIn}$ heusler alloy},\ }\href {https://doi.org/10.1103/PhysRevB.101.014443} {\bibfield  {journal} {\bibinfo  {journal} {Phys. Rev. B}\ }\textbf {\bibinfo {volume} {101}},\ \bibinfo {pages} {014443} (\bibinfo {year} {2020})}\BibitemShut {NoStop}%
\bibitem [{\citenamefont {Mulazzi}\ \emph {et~al.}(2008)\citenamefont {Mulazzi}, \citenamefont {Chainani}, \citenamefont {Takata}, \citenamefont {Tanaka}, \citenamefont {Nishino}, \citenamefont {Tamasaku}, \citenamefont {Ishikawa}, \citenamefont {Takeuchi}, \citenamefont {Ishida}, \citenamefont {Senba}, \citenamefont {Ohashi},\ and\ \citenamefont {Shin}}]{Mulazzi_2008}%
  \BibitemOpen
  \bibfield  {author} {\bibinfo {author} {\bibfnamefont {M.}~\bibnamefont {Mulazzi}}, \bibinfo {author} {\bibfnamefont {A.}~\bibnamefont {Chainani}}, \bibinfo {author} {\bibfnamefont {Y.}~\bibnamefont {Takata}}, \bibinfo {author} {\bibfnamefont {Y.}~\bibnamefont {Tanaka}}, \bibinfo {author} {\bibfnamefont {Y.}~\bibnamefont {Nishino}}, \bibinfo {author} {\bibfnamefont {K.}~\bibnamefont {Tamasaku}}, \bibinfo {author} {\bibfnamefont {T.}~\bibnamefont {Ishikawa}}, \bibinfo {author} {\bibfnamefont {T.}~\bibnamefont {Takeuchi}}, \bibinfo {author} {\bibfnamefont {Y.}~\bibnamefont {Ishida}}, \bibinfo {author} {\bibfnamefont {Y.}~\bibnamefont {Senba}}, \bibinfo {author} {\bibfnamefont {H.}~\bibnamefont {Ohashi}},\ and\ \bibinfo {author} {\bibfnamefont {S.}~\bibnamefont {Shin}},\ }\bibfield  {title} {\bibinfo {title} {Temperature dependence of the exchange stiffness in {{FePd}}(001) thin films: Deviation from the empirical law $a(t)\ensuremath{\propto}{M}_{S}^{2}$ at intermediate temperatures},\ }\href
  {https://doi.org/10.1103/PhysRevB.77.224425} {\bibfield  {journal} {\bibinfo  {journal} {Phys. Rev. B}\ }\textbf {\bibinfo {volume} {77}},\ \bibinfo {pages} {224425} (\bibinfo {year} {2008})}\BibitemShut {NoStop}%
\bibitem [{\citenamefont {Moreno}\ \emph {et~al.}(2025)\citenamefont {Moreno}, \citenamefont {Bercoff}, \citenamefont {Atxitia}, \citenamefont {Evans},\ and\ \citenamefont {Chubykalo-Fesenko}}]{Moreno2024TempFerri}%
  \BibitemOpen
  \bibfield  {author} {\bibinfo {author} {\bibfnamefont {R.}~\bibnamefont {Moreno}}, \bibinfo {author} {\bibfnamefont {P.~G.}\ \bibnamefont {Bercoff}}, \bibinfo {author} {\bibfnamefont {U.}~\bibnamefont {Atxitia}}, \bibinfo {author} {\bibfnamefont {R.~F.~L.}\ \bibnamefont {Evans}},\ and\ \bibinfo {author} {\bibfnamefont {O.}~\bibnamefont {Chubykalo-Fesenko}},\ }\bibfield  {title} {\bibinfo {title} {Temperature dependence of exchange stiffness and energy barrier in compensated ferrimagnets},\ }\href {https://doi.org/10.1103/PhysRevB.111.184416} {\bibfield  {journal} {\bibinfo  {journal} {Physical Review B}\ }\textbf {\bibinfo {volume} {111}},\ \bibinfo {pages} {184416} (\bibinfo {year} {2025})}\BibitemShut {NoStop}%
\bibitem [{\citenamefont {Hirst}\ \emph {et~al.}(2022)\citenamefont {Hirst}, \citenamefont {Atxitia}, \citenamefont {Ruta}, \citenamefont {Jackson}, \citenamefont {Petit},\ and\ \citenamefont {Ostler}}]{Hirst2022TempMn2Au}%
  \BibitemOpen
  \bibfield  {author} {\bibinfo {author} {\bibfnamefont {J.}~\bibnamefont {Hirst}}, \bibinfo {author} {\bibfnamefont {U.}~\bibnamefont {Atxitia}}, \bibinfo {author} {\bibfnamefont {S.}~\bibnamefont {Ruta}}, \bibinfo {author} {\bibfnamefont {J.}~\bibnamefont {Jackson}}, \bibinfo {author} {\bibfnamefont {L.}~\bibnamefont {Petit}},\ and\ \bibinfo {author} {\bibfnamefont {T.}~\bibnamefont {Ostler}},\ }\bibfield  {title} {\bibinfo {title} {Temperature-dependent micromagnetic model of the antiferromagnet {{Mn}}$_2${{Au}}: A multiscale approach},\ }\href {https://doi.org/10.1103/PhysRevB.106.094402} {\bibfield  {journal} {\bibinfo  {journal} {Physical Review B}\ }\textbf {\bibinfo {volume} {106}},\ \bibinfo {pages} {094402} (\bibinfo {year} {2022})}\BibitemShut {NoStop}%
\bibitem [{\citenamefont {Rozsa}\ and\ \citenamefont {Atxitia}(2023)}]{Rozsa2022TempAFerro}%
  \BibitemOpen
  \bibfield  {author} {\bibinfo {author} {\bibfnamefont {L.}~\bibnamefont {Rozsa}}\ and\ \bibinfo {author} {\bibfnamefont {U.}~\bibnamefont {Atxitia}},\ }\bibfield  {title} {\bibinfo {title} {Temperature dependence of spin-modeled parameters in antiferromagnets},\ }\href {https://doi.org/10.1103/PhysRevResearch.5.023139} {\bibfield  {journal} {\bibinfo  {journal} {Physical Review Research}\ }\textbf {\bibinfo {volume} {5}},\ \bibinfo {pages} {023139} (\bibinfo {year} {2023})}\BibitemShut {NoStop}%
\bibitem [{\citenamefont {Mikuni}\ \emph {et~al.}(2025)\citenamefont {Mikuni}, \citenamefont {Kuramoto}, \citenamefont {Fujii}, \citenamefont {Koreeda}, \citenamefont {Parchenko}, \citenamefont {Stupakiewicz},\ and\ \citenamefont {Satoh}}]{Mikuni2025FMRAFerro}%
  \BibitemOpen
  \bibfield  {author} {\bibinfo {author} {\bibfnamefont {K.}~\bibnamefont {Mikuni}}, \bibinfo {author} {\bibfnamefont {T.}~\bibnamefont {Kuramoto}}, \bibinfo {author} {\bibfnamefont {Y.}~\bibnamefont {Fujii}}, \bibinfo {author} {\bibfnamefont {A.}~\bibnamefont {Koreeda}}, \bibinfo {author} {\bibfnamefont {S.}~\bibnamefont {Parchenko}}, \bibinfo {author} {\bibfnamefont {A.}~\bibnamefont {Stupakiewicz}},\ and\ \bibinfo {author} {\bibfnamefont {T.}~\bibnamefont {Satoh}},\ }\bibfield  {title} {\bibinfo {title} {Magnetic resonance frequency of two-sublattice ferrimagnet with magnetic compensation temperature},\ }\href {https://doi.org/10.7566/JPSJ.94.111001} {\bibfield  {journal} {\bibinfo  {journal} {Journal of the Physical Society Japan}\ }\textbf {\bibinfo {volume} {94}},\ \bibinfo {pages} {111001} (\bibinfo {year} {2025})}\BibitemShut {NoStop}%
\bibitem [{\citenamefont {Kikkawa}\ \emph {et~al.}(2015)\citenamefont {Kikkawa}, \citenamefont {Uchida}, \citenamefont {Daimon}, \citenamefont {Qiu}, \citenamefont {Shiomi},\ and\ \citenamefont {Saitoh}}]{kikkawa2015critical}%
  \BibitemOpen
  \bibfield  {author} {\bibinfo {author} {\bibfnamefont {T.}~\bibnamefont {Kikkawa}}, \bibinfo {author} {\bibfnamefont {K.-i.}\ \bibnamefont {Uchida}}, \bibinfo {author} {\bibfnamefont {S.}~\bibnamefont {Daimon}}, \bibinfo {author} {\bibfnamefont {Z.}~\bibnamefont {Qiu}}, \bibinfo {author} {\bibfnamefont {Y.}~\bibnamefont {Shiomi}},\ and\ \bibinfo {author} {\bibfnamefont {E.}~\bibnamefont {Saitoh}},\ }\bibfield  {title} {\bibinfo {title} {Critical suppression of spin seebeck effect by magnetic fields},\ }\href {https://doi.org/10.1103/PhysRevB.92.064413} {\bibfield  {journal} {\bibinfo  {journal} {Physical Review B}\ }\textbf {\bibinfo {volume} {92}},\ \bibinfo {pages} {064413} (\bibinfo {year} {2015})}\BibitemShut {NoStop}%
\bibitem [{\citenamefont {Jin}\ \emph {et~al.}(2015)\citenamefont {Jin}, \citenamefont {Boona}, \citenamefont {Yang}, \citenamefont {Myers},\ and\ \citenamefont {Heremans}}]{jin2015effect}%
  \BibitemOpen
  \bibfield  {author} {\bibinfo {author} {\bibfnamefont {H.}~\bibnamefont {Jin}}, \bibinfo {author} {\bibfnamefont {S.~R.}\ \bibnamefont {Boona}}, \bibinfo {author} {\bibfnamefont {Z.}~\bibnamefont {Yang}}, \bibinfo {author} {\bibfnamefont {R.~C.}\ \bibnamefont {Myers}},\ and\ \bibinfo {author} {\bibfnamefont {J.~P.}\ \bibnamefont {Heremans}},\ }\bibfield  {title} {\bibinfo {title} {Effect of the magnon dispersion on the longitudinal spin seebeck effect in yttrium iron garnets},\ }\href {https://doi.org/10.1103/PhysRevB.92.054436} {\bibfield  {journal} {\bibinfo  {journal} {Physical Review B}\ }\textbf {\bibinfo {volume} {92}},\ \bibinfo {pages} {054436} (\bibinfo {year} {2015})}\BibitemShut {NoStop}%
\bibitem [{\citenamefont {Rezende}\ \emph {et~al.}(2014)\citenamefont {Rezende}, \citenamefont {Rodr{\'\i}guez-Su{\'a}rez}, \citenamefont {Cunha}, \citenamefont {Rodrigues}, \citenamefont {Machado}, \citenamefont {Fonseca~Guerra}, \citenamefont {Lopez~Ortiz},\ and\ \citenamefont {Azevedo}}]{rezende2014magnon}%
  \BibitemOpen
  \bibfield  {author} {\bibinfo {author} {\bibfnamefont {S.}~\bibnamefont {Rezende}}, \bibinfo {author} {\bibfnamefont {R.}~\bibnamefont {Rodr{\'\i}guez-Su{\'a}rez}}, \bibinfo {author} {\bibfnamefont {R.}~\bibnamefont {Cunha}}, \bibinfo {author} {\bibfnamefont {A.}~\bibnamefont {Rodrigues}}, \bibinfo {author} {\bibfnamefont {F.}~\bibnamefont {Machado}}, \bibinfo {author} {\bibfnamefont {G.}~\bibnamefont {Fonseca~Guerra}}, \bibinfo {author} {\bibfnamefont {J.}~\bibnamefont {Lopez~Ortiz}},\ and\ \bibinfo {author} {\bibfnamefont {A.}~\bibnamefont {Azevedo}},\ }\bibfield  {title} {\bibinfo {title} {Magnon spin-current theory for the longitudinal spin-seebeck effect},\ }\href {https://doi.org/10.1103/PhysRevB.89.014416} {\bibfield  {journal} {\bibinfo  {journal} {Physical Review B}\ }\textbf {\bibinfo {volume} {89}},\ \bibinfo {pages} {014416} (\bibinfo {year} {2014})}\BibitemShut {NoStop}%
\bibitem [{\citenamefont {Xiao}\ \emph {et~al.}(2010)\citenamefont {Xiao}, \citenamefont {Bauer}, \citenamefont {Uchida}, \citenamefont {Saitoh},\ and\ \citenamefont {Maekawa}}]{xiao2010theory}%
  \BibitemOpen
  \bibfield  {author} {\bibinfo {author} {\bibfnamefont {J.}~\bibnamefont {Xiao}}, \bibinfo {author} {\bibfnamefont {G.~E.}\ \bibnamefont {Bauer}}, \bibinfo {author} {\bibfnamefont {K.-c.}\ \bibnamefont {Uchida}}, \bibinfo {author} {\bibfnamefont {E.}~\bibnamefont {Saitoh}},\ and\ \bibinfo {author} {\bibfnamefont {S.}~\bibnamefont {Maekawa}},\ }\bibfield  {title} {\bibinfo {title} {Theory of magnon-driven spin seebeck effect},\ }\href {https://doi.org/10.1103/PhysRevB.81.214418} {\bibfield  {journal} {\bibinfo  {journal} {Physical Review B—Condensed Matter and Materials Physics}\ }\textbf {\bibinfo {volume} {81}},\ \bibinfo {pages} {214418} (\bibinfo {year} {2010})}\BibitemShut {NoStop}%
\bibitem [{\citenamefont {Mihalceanu}\ \emph {et~al.}(2018)\citenamefont {Mihalceanu}, \citenamefont {Vasyuchka}, \citenamefont {Bozhko}, \citenamefont {Langner}, \citenamefont {Nechiporuk}, \citenamefont {Romanyuk}, \citenamefont {Hillebrands},\ and\ \citenamefont {Serga}}]{mihalceanu2018temperature}%
  \BibitemOpen
  \bibfield  {author} {\bibinfo {author} {\bibfnamefont {L.}~\bibnamefont {Mihalceanu}}, \bibinfo {author} {\bibfnamefont {V.~I.}\ \bibnamefont {Vasyuchka}}, \bibinfo {author} {\bibfnamefont {D.~A.}\ \bibnamefont {Bozhko}}, \bibinfo {author} {\bibfnamefont {T.}~\bibnamefont {Langner}}, \bibinfo {author} {\bibfnamefont {A.~Y.}\ \bibnamefont {Nechiporuk}}, \bibinfo {author} {\bibfnamefont {V.~F.}\ \bibnamefont {Romanyuk}}, \bibinfo {author} {\bibfnamefont {B.}~\bibnamefont {Hillebrands}},\ and\ \bibinfo {author} {\bibfnamefont {A.~A.}\ \bibnamefont {Serga}},\ }\bibfield  {title} {\bibinfo {title} {Temperature-dependent relaxation of dipole-exchange magnons in yttrium iron garnet films},\ }\href {https://doi.org/10.1103/PhysRevB.97.214405} {\bibfield  {journal} {\bibinfo  {journal} {Physical Review B}\ }\textbf {\bibinfo {volume} {97}},\ \bibinfo {pages} {214405} (\bibinfo {year} {2018})}\BibitemShut {NoStop}%
\end{thebibliography}%

\end{document}